\documentclass[prx,twocolumn,amsmath,amssymb,dvipdfmx]{revtex4}

\usepackage{color}
\usepackage{bm}
\usepackage{dcolumn}
\usepackage[dvipdfmx]{graphicx}
\usepackage{slashed}
\usepackage{amsmath}\usepackage{accents}
\usepackage{ulem}
\usepackage{mathdots}

\newcommand{\1}{\mbox{1}\hspace{-0.25em}\mbox{l}}

\newcommand{\tr}{{\rm tr}\,}

\begin{document}

\title{
Higher-order topological insulators in a magnetic field
}

\author{Yuria Otaki and Takahiro Fukui}
\affiliation{Department of Physics, Ibaraki University, Mito 310-8512, Japan}

\date{\today}

\begin{abstract}
Two-dimensional (2D) generalization of the Su-Schriffer-Heeger (SSH) model serves as a platform for
exploring higher-order topological insulators (HOTI). 
We investigate this model in a magnetic field which interpolates two  models studied so far
with zero flux and $\pi$ flux per plaquette.  
We show that in the Hofstadter butterfly there appears  a wide gap around the $\pi$ flux, 
which belongs to the same HOTI discovered by  Benalcazar-Bernevig-Hughes (BBH). 
It turns out that in a weak field regime HOTI could exist even within a small gap disconnected from 
the wider gap around $\pi$ flux. 
To characterize HOTI, we employ the entanglement polarization (eP) technique
which is useful even if the basic four bands split into many Landau levels under a magnetic field.
\end{abstract}

\pacs{
}

\maketitle

\section{Introduction}

Topological classification of matter is nowadays one of the fundamental methods to understand various phenomena 
in condensed matter physics \cite{Kane:2005aa,Qi:2008aa,Schnyder:2008aa,Teo:2010fk,Hasan:2010fk,Qi:2011kx}.
In addition to time reversal, particle-hole, and chiral symmetries, crystalline point group symmetries enrich periodic tables
of topological insulators and superconductors \cite{Fu:2011aa,Morimoto:2013aa,Shiozaki:2014aa,Kruthoff:2017aa}.
Topological classification has also opened a new venue to explore topological phenomena 
in metamaterials  such as  phononic systems \cite{Kane:2013aa,Kariyado:2015aa}, 
photonic crystals \cite{Wang:2009aa,Khanikaev:2012aa}, and so on.
Experimentally, edge states associated with topological properties of bulk play a crucial role as observables.
This is the bulk-edge correspondence \cite{Hatsugai:1993fk}.

The recent discovery of higher-order topological insulators (HOTI)
\cite{Benalcazar:2017aa,Benalcazar:2017ab,Schindler:2018aa, Hayashi:2018aa,Hashimoto:2017aa} 
has led us to a renewed interest 
in the bulk-edge correspondence. 
For conventional topological insulators, bulk topological invariants are directly related to gapless boundary states
\cite{Hatsugai:1993fk,Kane:2005aa}.
In HOTI, on the other hand, both of them seem trivial, that is, bulk topological invariants vanish and boundary states are gapped out.
Nevertheless, higher-order boundary states such as corner or hinge states show up. These states are guaranteed by 
higher-order topological invariants of the bulk, e.g.,  one-dimensional (1D) Berry-Zak phases in two- and higher-dimensional systems.
This implies that there is still ``higher-order" bulk-edge correspondence \cite{1908.00011}.
HOTI have been attracting much current interest
\cite{Langbehn:2017aa,Song:2017aa,Ezawa:2018aa,Ezawa:2018ab,Liu:2017aa,Khalaf:2018cr,Matsugatani:2018aa,Fukui:2018aa,
Calugaru:2019aa},
and observed experimentally in various metamaterial systems
\cite{Imhof:2018aa,Zhang:2019ab,Ota:2019aa}.

\begin{figure}[htb]
\begin{center}
\includegraphics[width=.9\linewidth]{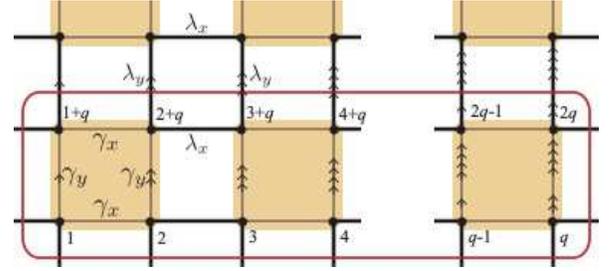}
\caption{
The lattice on which the model is defined. The thin and thick lines stand for the alternating 
bond strength. The SSH unit cell is colored orange. 
The $j$ arrow shows the Peierls phase factor $e^{i(j-\frac{1}{2})\phi}$
for a uniform flux $\phi=2\pi p/q$ per plaquette.
The magnetic unit cell is surrounded by the red square.
}
\label{f:lat}
\end{center}
\end{figure}

One of the typical models for HOTI
is a two-dimensional (2D) generalization \cite{Benalcazar:2017aa,Benalcazar:2017ab, Liu:2017aa}
of the SSH model \cite{Su:1979aa}.
Consider a tight-binding model on the square lattice with nearest-neighbor hoppings only, 
as shown in Fig. \ref{f:lat} ($\phi=0$).
Then, the model is a simple decoupled SSH model 
$H=h_{{\rm SSH},x}\otimes 1+1\otimes h_{{\rm SSH},y}$ in the momentum space, 
where $h_{{\rm SSH},j}$ 
stands for the 1D SSH Hamiltonian toward the $j$ direction 
\cite{Benalcazar:2017aa,Benalcazar:2017ab, Liu:2017aa}. 
Therefore, it is obvious that the model shows corner states
as the edge states of the 1D SSH models. 
Remember that 
these edge states  are
protected by the chiral symmetry of each chain $\{h_{{\rm SSH},j},\sigma\}=0$, 
where $\sigma$ is a certain matrix depending on the representation.
Its topological invariant is the  quantized polarization (Berry-Zak phase) \cite{Ryu:2002fk}.
Therefore, the corner states of this model are also ensured by the
quantized polarizations for both directions \cite{Liu:2017aa}.
On the other hand, BBH introduced $\pi$-flux per plaquette to this model, as shown in Fig. \ref{f:lat} with $\phi=\pi$.
In this case, the Hamiltonian can be written as 
$H=h_{{\rm SSH},x}\otimes 1+\sigma\otimes h_{{\rm SSH},y}$ in the momentum space.
Because of the anti-commutability of the $x$ and $y$ sectors, the model becomes gapful.
Its  ground state is characterized by the topological quadrupole moment
\cite{Benalcazar:2017aa,Benalcazar:2017ab}, 
or the product formula of topological invariants in the mathematical context \cite{Hayashi:2018aa}.

These two models studied so far are the 2D  SSH model with 0-flux and $\pi$-flux.
With $0$-flux, the half-filled ground state is basically gapless and the corner states are embedded in the bulk spectrum,
whereas a magnetic field giving $\pi$-flux per plaquette is too strong to realize in experiments.
Therefore, the quest for the possibility of HOTI in an arbitrary magnetic field is not only a theoretical interest 
but also 
extends the possibility of experimental observations of HOTI in real materials.

In this paper, we generalize these two models
by introducing  generic magnetic flux $\phi$
which interpolates the simple 2D SSH model at $\phi=0$ and the BBH model at $\phi=\pi$.
We show that in the Hofstadter butterfly, there appear several gaps at half-filling separated by gap-closings. 
To investigate the topological properties of these gapped states, we use the entanglement technique 
developed by the authors Refs. \cite{Fukui:2014qv,Fukui:2015fk} and successfully applied to the BBH model \cite{Fukui:2018aa}.
We show that this method is simple enough to compute eP characterizing HOTI
even if the bands of the system split into many Landau levels.
We argue that there appear half-filled HOTI even in a weak field regime which may be accessible by experiments.
Previously, another model interpolating those with 0-flux and $\pi$-flux was proposed in 
\cite{1812.06990}, which introduced a local flux with zero mean. In contrast to this model, our model in the present paper
includes a uniform magnetic flux, and it seems more feasible in experiments.  

This paper is organized as follows.
In the next section, we present the model Hamiltonian and show several numerical results including Hofstadter buttery 
spectra and corner states.
In Sec. \ref{s:ep}, we introduce eP as topological invariants characterizing HOTI in a magnetic field, and show 
several numerical examples of computed eP.
In Sec. \ref{s:ex}, we briefly discuss the experimental feasibility of the HOTI in a weak magnetic field regime,
introducing symmetry-breaking potentials.
In Sec. \ref{s:sum}, we give the summary and discussion.

\section{2D SHH model in a magnetic field}\label{s:modelt}

In this section, we introduce the 2D SSH model Hamiltonian in a uniform magnetic field.
In Sec. \ref{s:model}, we present the matrix elements of the Bloch Hamiltonian, and discuss the
symmetries of the model. The fundamental reflection symmetries are modified due to the magnetic flux.
These symmetries play a crucial role in the quantization of the eP, as will be discussed in Sec. \ref{s:ep}.
We next discuss the possibility of the HOTI realized in the Hofstadter butterfly spectra in Sec. \ref{s:but}, 
and show several examples of corner states in Sec. \ref{s:cor}.
These are indeed characterized by the nontrivial eP, as will be demonstrated in Sec. \ref{s:ep}.

\subsection{Model}\label{s:model}
The model is defined on the square lattice with the nearest neighbor hopping,
\begin{alignat}1
H(\phi)=\sum_{\langle i,j\rangle} t_{i,j}c_i^\dagger c_j+\mbox{H.c.}=\sum_k\bm c_k^\dagger {\cal H}(k,\phi)\bm c_k,
\label{OriHam}
\end{alignat}
where $t_{i,j}$ is given by $t_{j+\hat x,j}=t_x$ and $t_{j+\hat y,j}=e^{i(j_x-1/2)\phi}t_y$. Here 
the real parameters $t_x$ and $t_y$ are $t_x=\gamma_x,\,t_y=\gamma_y$ within unit cells, 
whereas $t_x=\lambda_x,\,t_y=\lambda_y$
between unit cells, as illustrated in Fig. \ref{f:lat}.
We basically set $\gamma_x=\gamma_y\equiv\gamma$ and $\lambda_x=\lambda_y\equiv\lambda\,(=1)$.
A uniform flux $\phi=2\pi p/q$ per plaquette is introduced in the Landau gauge.
For such a gauge fixing and choice of the magnetic unit cell,  
it may be natural to choose the Brillouin zone as $|k_x|\le\pi/q$ and $|k_y|\le \pi/2$. 

\subsubsection{Hamiltonian in the momentum representation}

The Hamiltonian in the momentum space in Eq. (\ref{OriHam})
is given by
\begin{alignat}1
{\cal H}(k,\phi)=
\begin{pmatrix}
{\cal H}_x(k_x)&{\cal H}_y(k_y,\phi)\\
{\cal H}_y^\dagger(k_y,\phi)&{\cal H}_x(k_x)
\end{pmatrix},
\end{alignat}
where ${\cal H}_x(k_x)$ and ${\cal H}_y(k_y,\phi)$ are $q\times q$ matrices associated with the hopping 
toward the $x$ and $y$ directions, respectively. They are explicitly given by
\begin{alignat}1
&{\cal H}_x(k_x)=
\left(
\begin{array}{ccccccc}
0&\gamma_x&&&&&e^{-iqk_x}\lambda_x\\
\gamma_x&0&\lambda_x&&&&\\
&\lambda_x&0&\gamma_x&&&\\
&&\gamma_x&&&&\\
&&&&\ddots&&\\
&&&&&&\gamma_x\\
e^{iqk_x}\lambda_x&&&&&\gamma_x&0
\end{array}
\right),
\nonumber\\
&{\cal H}_{y}(k_y,\phi)=
\mbox{diag}\left(\cdots,\underbrace{h_j(k_y,\phi)}_{j{\rm th}},\cdots \right),
\nonumber\\
&\quad h_j(k_y,\phi)\equiv\gamma_ye^{i\left(j-\frac{1}{2}\right)\phi}+\lambda_ye^{-2ik_y-i\left(j-\frac{1}{2}\right)\phi}.
\end{alignat}
Note that ${\cal H}_x^*(k_x)={\cal H}_x(-k_x)$, and ${\cal H}_y^\dagger(k_y,\phi)={\cal H}_y^*(k_y,\phi)={\cal H}_y(-k_y,-\phi)$.

\begin{figure*}[htb]
\begin{center}
\begin{tabular}{ccccccc}
\includegraphics[width=.13\linewidth]{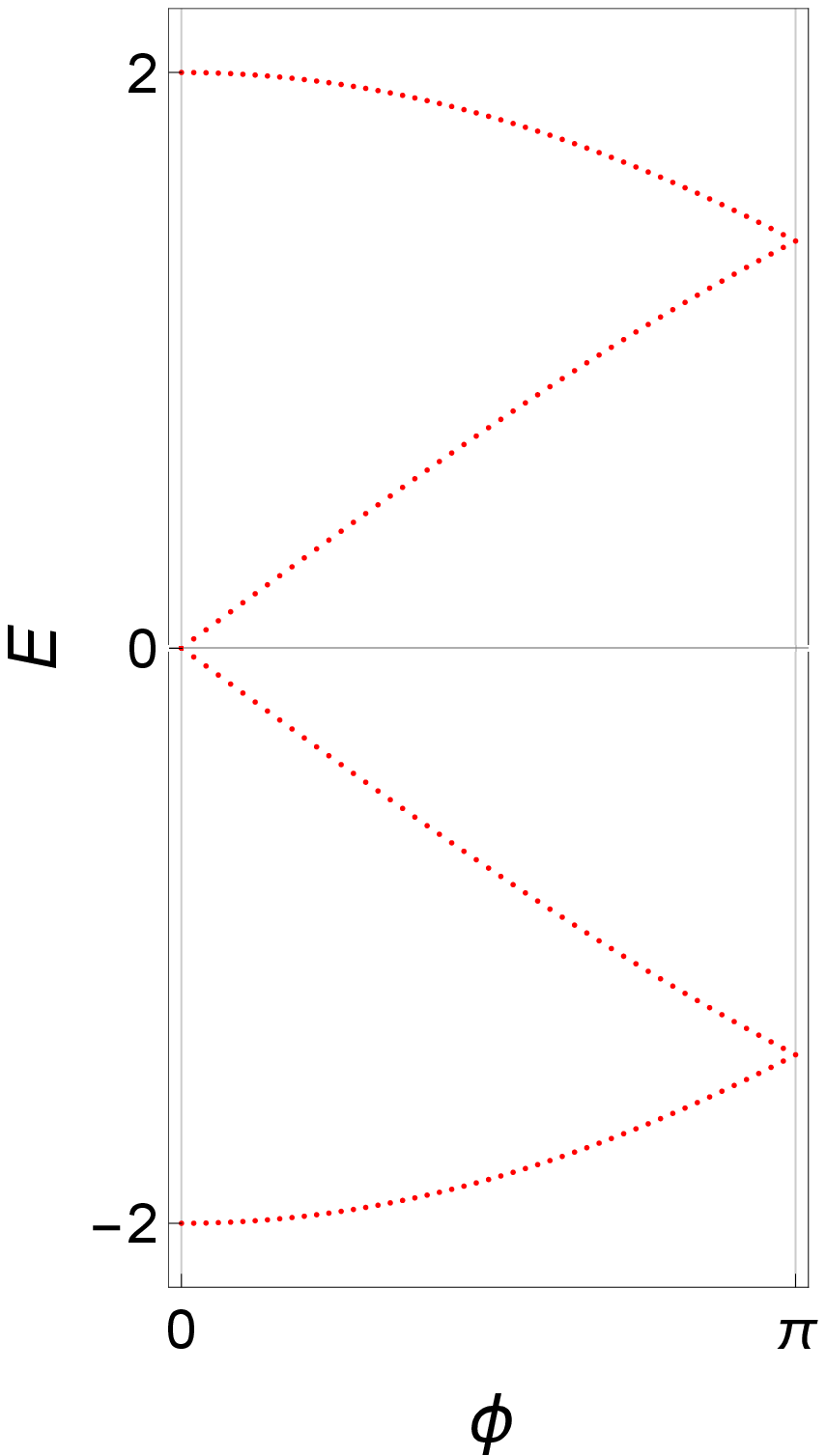}
&
\includegraphics[width=.13\linewidth]{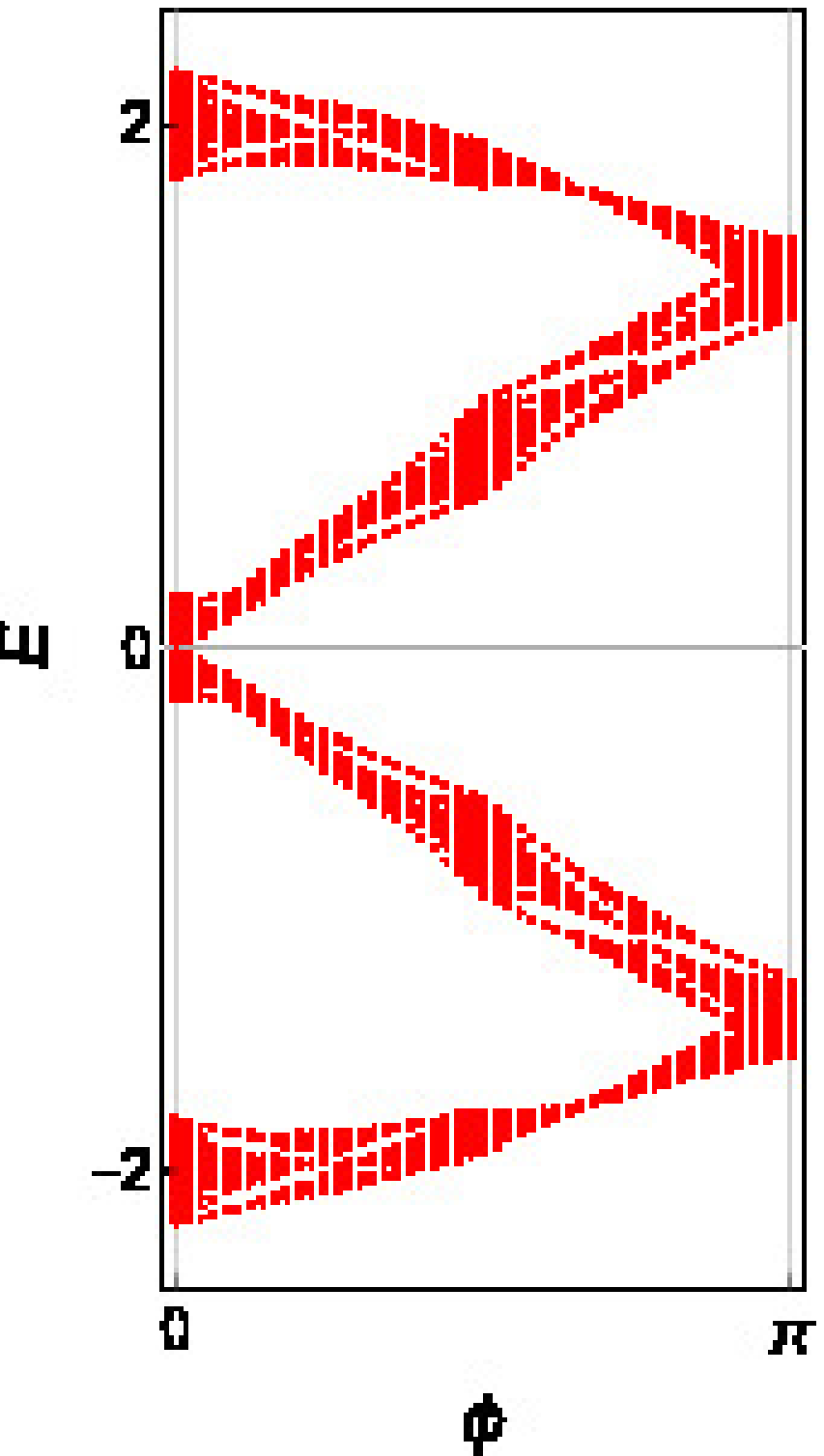}
&
\includegraphics[width=.13\linewidth]{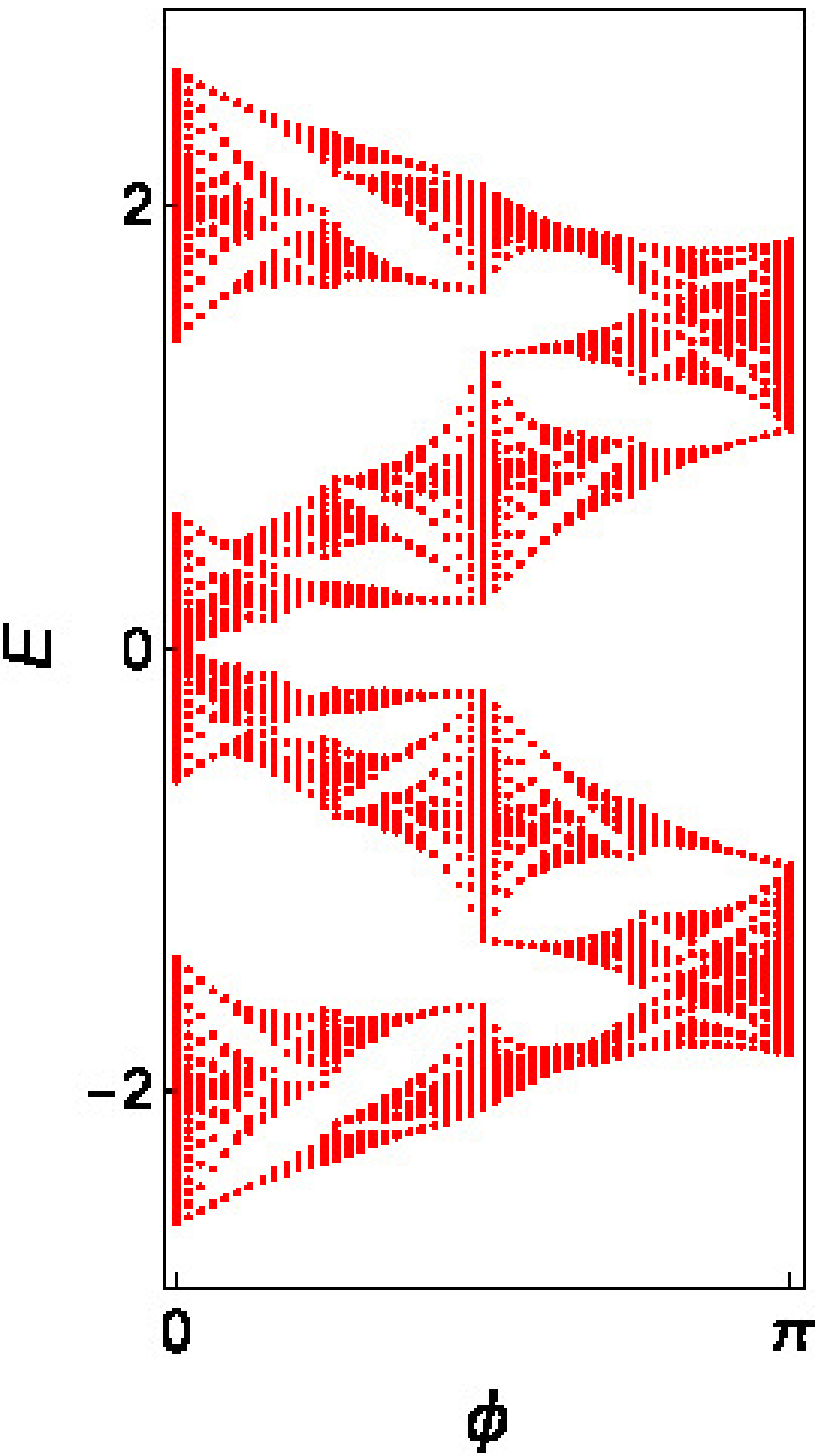}
&
\includegraphics[width=.13\linewidth]{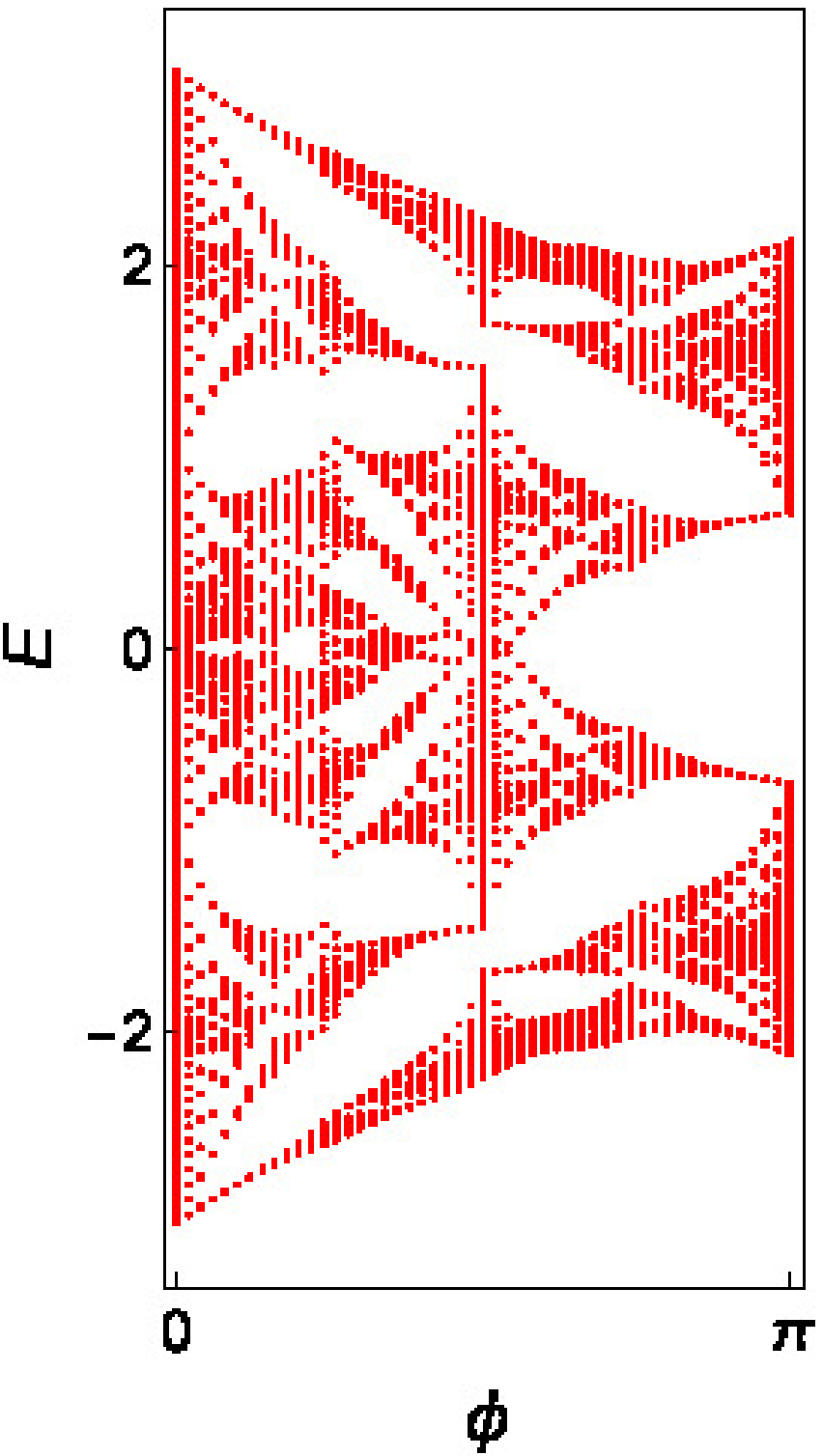}
&
\includegraphics[width=.13\linewidth]{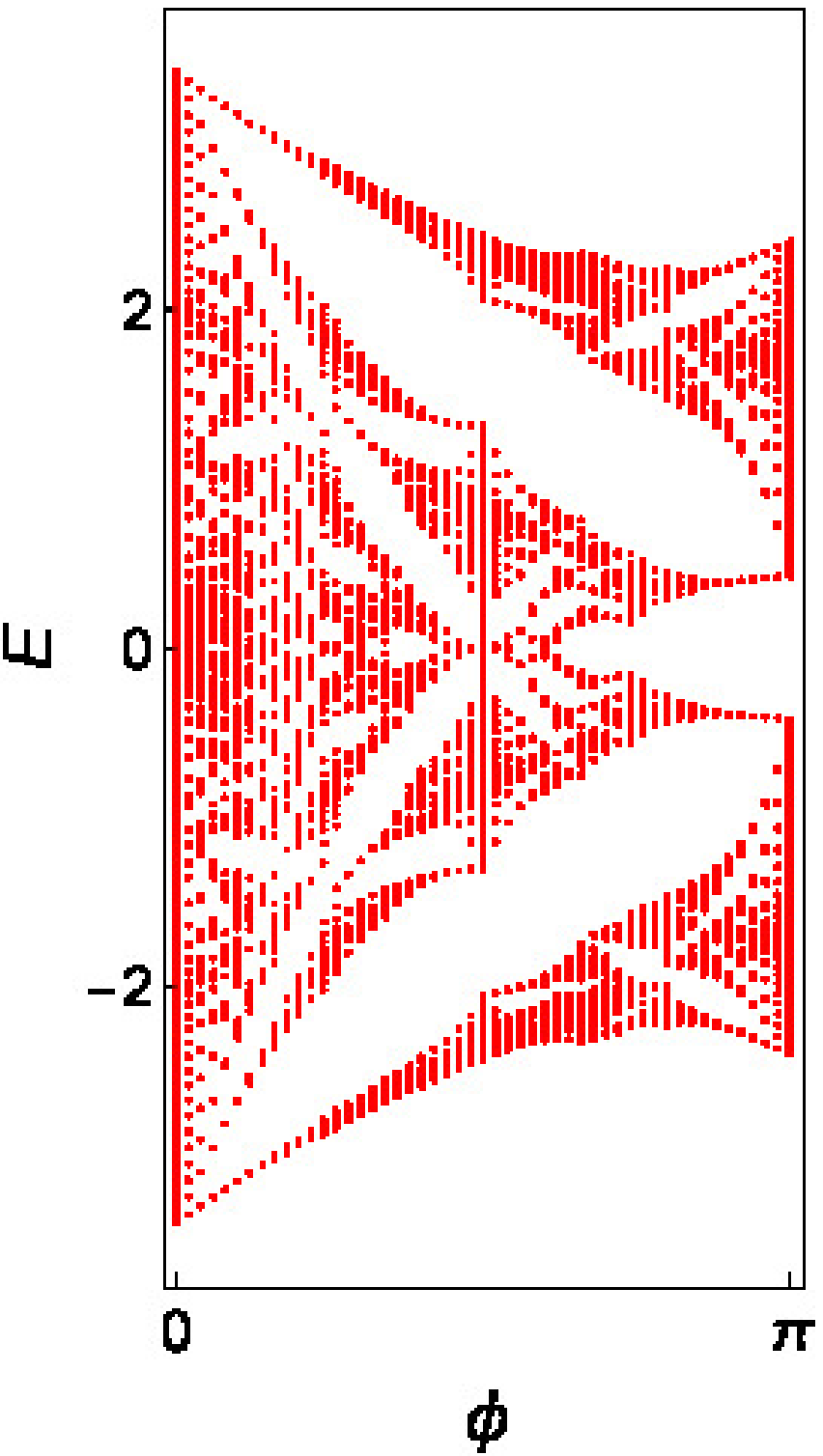}
&
\includegraphics[width=.13\linewidth]{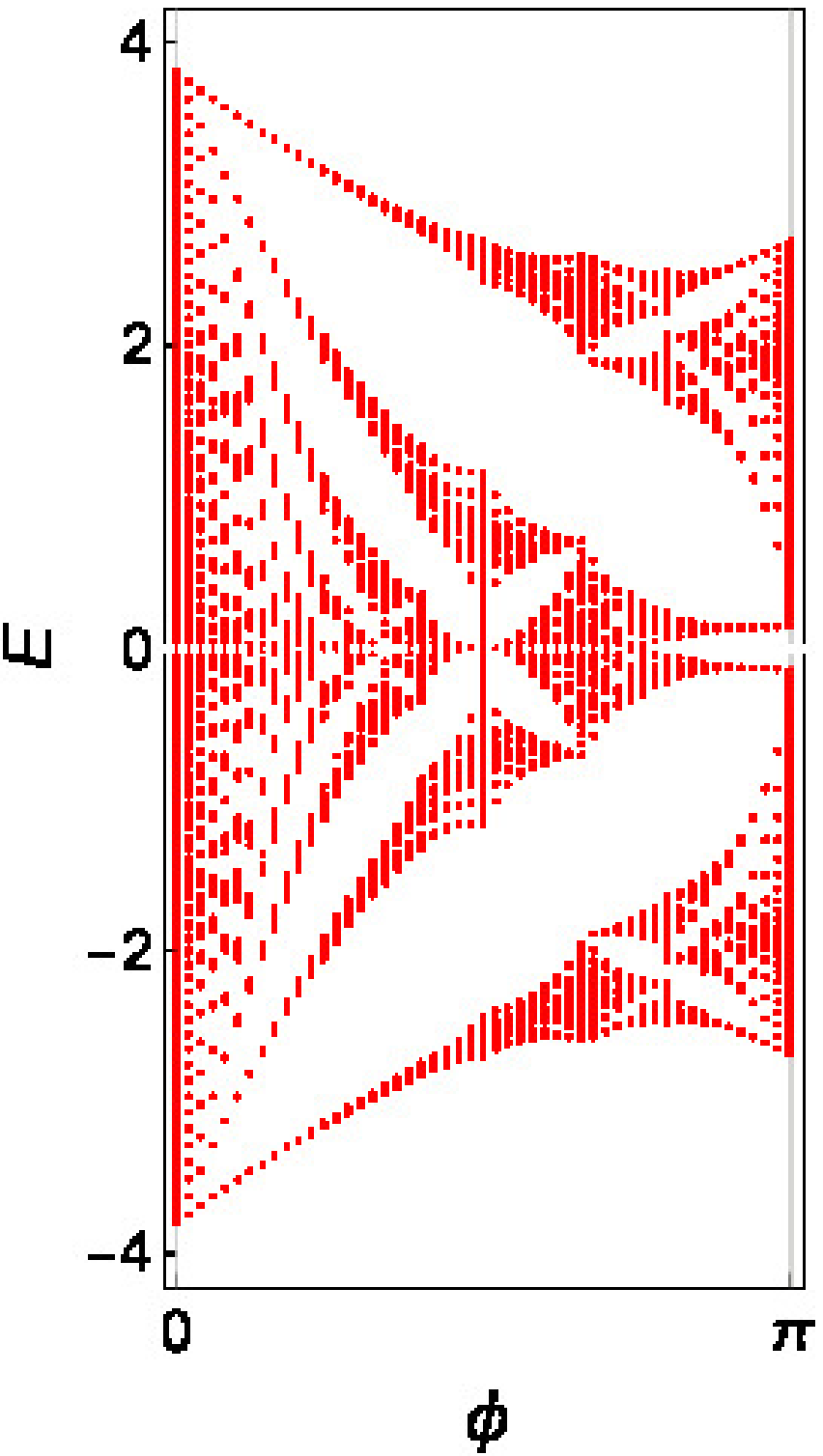}
&
\includegraphics[width=.13\linewidth]{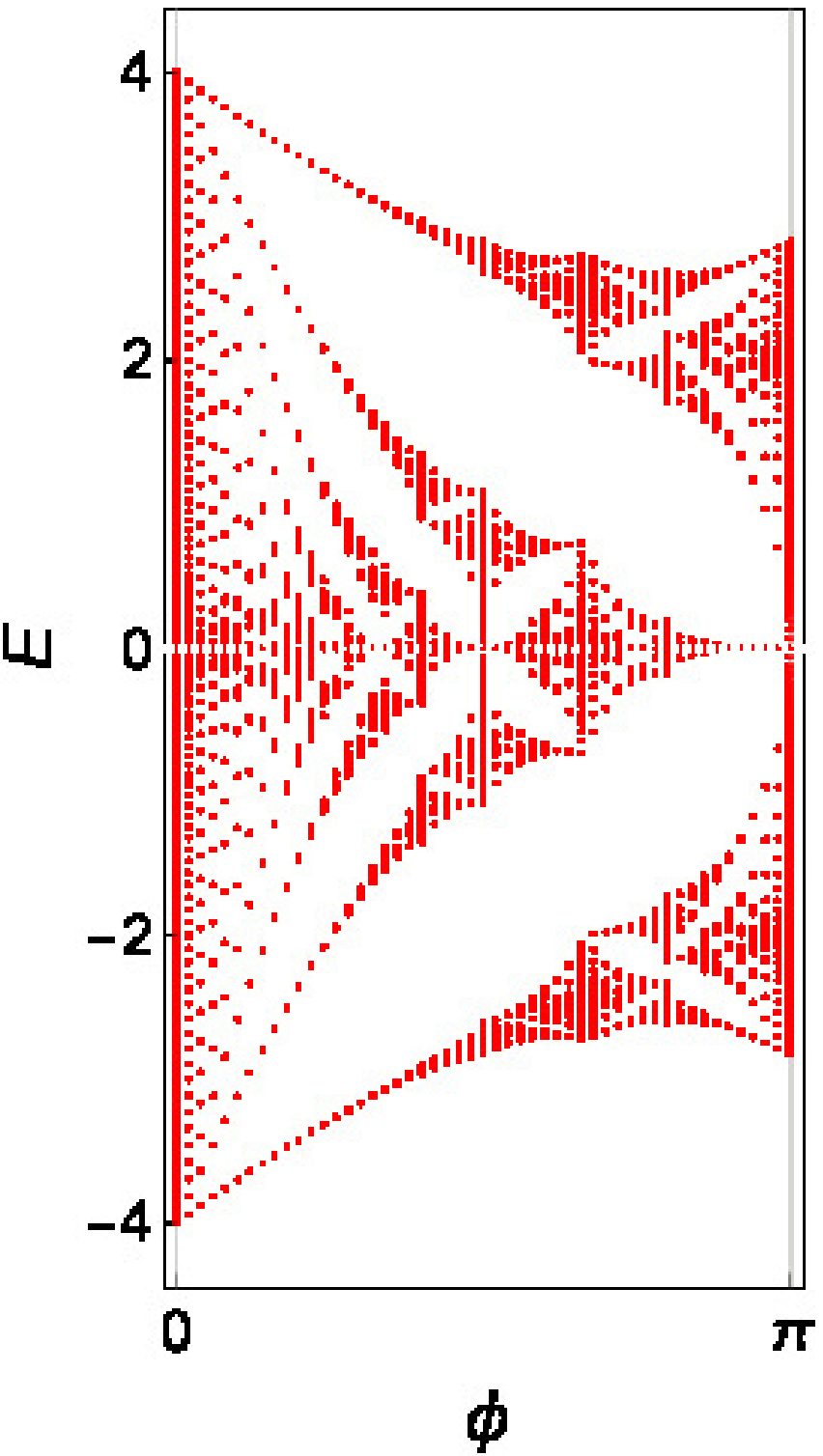}
\end{tabular}
\caption{
Hofstadter butterfly spectra for $\gamma=0$, $\gamma=0.1$, $\gamma=0.3$, $\gamma=0.5$, $\gamma=0.7$,
$\gamma=0.9$, and $\gamma=1$.  
}
\label{f:series_Hof}
\end{center}
\end{figure*}

\subsubsection{Symmetries}

As discussed in Refs. \cite{Benalcazar:2017aa,Benalcazar:2017ab}, reflection symmetries play a 
crucial role in the quantization of the topological quadrupole moment. 
In the present model, however, they change not only the sign of the momentum but also
the sign of the magnetic field such that
\begin{alignat}1
&M_x{\cal H}(k_x,k_y,\phi)M_x^{-1}={\cal H}(-k_x,k_y,-\phi),
\nonumber\\
&M_y{\cal H}(k_x,k_y,\phi)M_y^{-1}={\cal H}(k_x,-k_y,-\phi).
\label{OriTraLaw}
\end{alignat}
To show above, note that
the reflection in the $x$ direction induces the exchange of sites in the magnetic unit cell, $j\,(+q)\rightarrow q-j+1\,(+q)$.
This can be represented by the use of the following $q\times q$ matrix
\begin{alignat}1
\Sigma=
\begin{pmatrix}
&&&&&1\\
&&&1&&\\
&\iddots&&&&\\
1&&&&&
\end{pmatrix},
\end{alignat}
which induces the following transformations:
\begin{alignat}1
\Sigma{\cal H}_x(k_x)\Sigma^{-1}&={\cal H}_x(-k_x),
\nonumber\\
\Sigma{\cal H}_y(k_y,\phi)\Sigma^{-1}&=
\mbox{diag}\left(\cdots,\underbrace{h_{q-j+1}(k_y,\phi)}_{=h_j(k_y,-\phi)},\cdots \right)
\nonumber\\
&={\cal H}_y(k_y,-\phi).
\end{alignat}
Therefore, we can define $M_x$ by 
\begin{alignat}1
M_x=\begin{pmatrix}\Sigma&\\&\Sigma\end{pmatrix}.
\end{alignat}
This leads to the transformation law with respect to the $x$-reflection in Eq. (\ref{OriTraLaw}).
The reflection in the $y$ direction induces $j\leftrightarrow j+q$. Therefore, the following $M_y$,
\begin{alignat}1
M_y=\begin{pmatrix}&\1\\\1&\end{pmatrix},
\end{alignat}
where $\1$ stands for the $q\times q$ unit matrix,
gives the reflection law  with respect to the $y$ direction in Eq. (\ref{OriTraLaw}).

On the other hand, under time reversal $T=K$, where $K$ stands for the complex conjugation, 
the transformation law of ${\cal H}(k,\phi)$ reads
\begin{alignat}1
&T{\cal H}(k_x,k_y,\phi)T^{-1}={\cal H}(-k_x,-k_y,-\phi).
\label{TimRev}
\end{alignat}
Combining Eq. (\ref{OriTraLaw}) with Eq. (\ref{TimRev}),
we can define anti-unitary reflection symmetries $\tilde M_j=M_jT$ ($j=x,y$), 
under which the Hamiltonian transforms as 
\begin{alignat}1
&\tilde M_x{\cal H}(k_x,k_y,\phi)\tilde M_x={\cal H}(k_x,-k_y,\phi),
\nonumber\\
&\tilde M_y{\cal H}(k_x,k_y,\phi)\tilde M_y={\cal H}(-k_x,k_y,\phi).
\label{Sym}
\end{alignat} 
In what follows, we omit the dependence on $\phi$, considering the system with fixed $\phi$.

\subsection{Hofstadter butterfly spectra}\label{s:but}

We show in Fig. \ref{f:series_Hof} how Hofstadter butterfly spectra change between two limiting cases 
$\gamma=0$ and $\gamma=1$.
We observe the following:
(1) For small $\gamma$, the spectrum show a large gap at half-filling except for  $\phi\sim0$,
implying that 
the half-filled ground states for any finite flux
may be adiabatically deformed to the BBH ground state, and hence 
the HOTI phase seems robust against magnetic fields. 
Thus, the decoupled SSH model at $\phi=0$
opens a gap immediately if a small magnetic field is applied, 
and these gapped ground states would be in the same class as the BBH model.
\begin{figure}[htb]
\begin{center}
\begin{tabular}{cc}
\includegraphics[width=.45\linewidth]{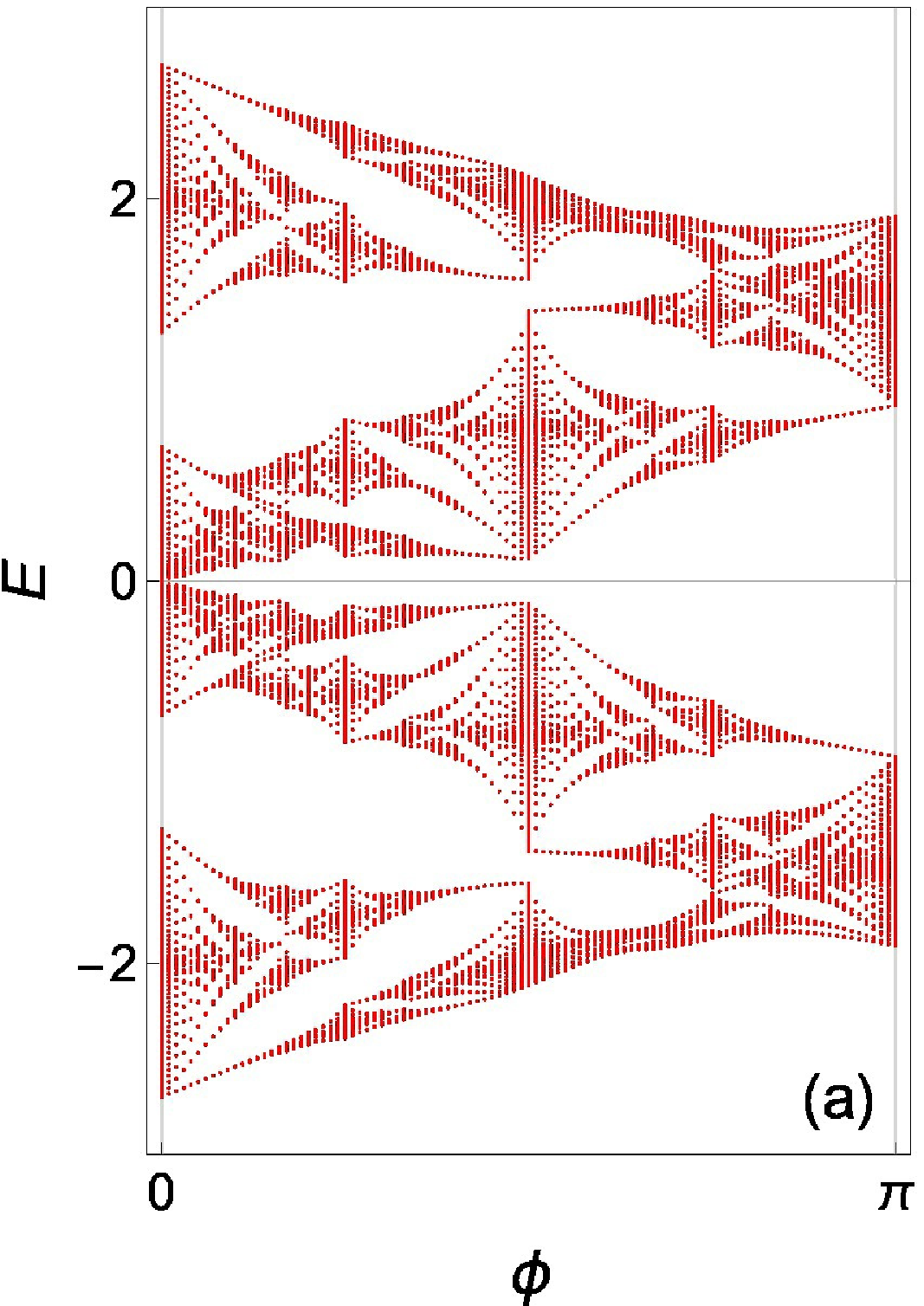}
&
\includegraphics[width=.45\linewidth]{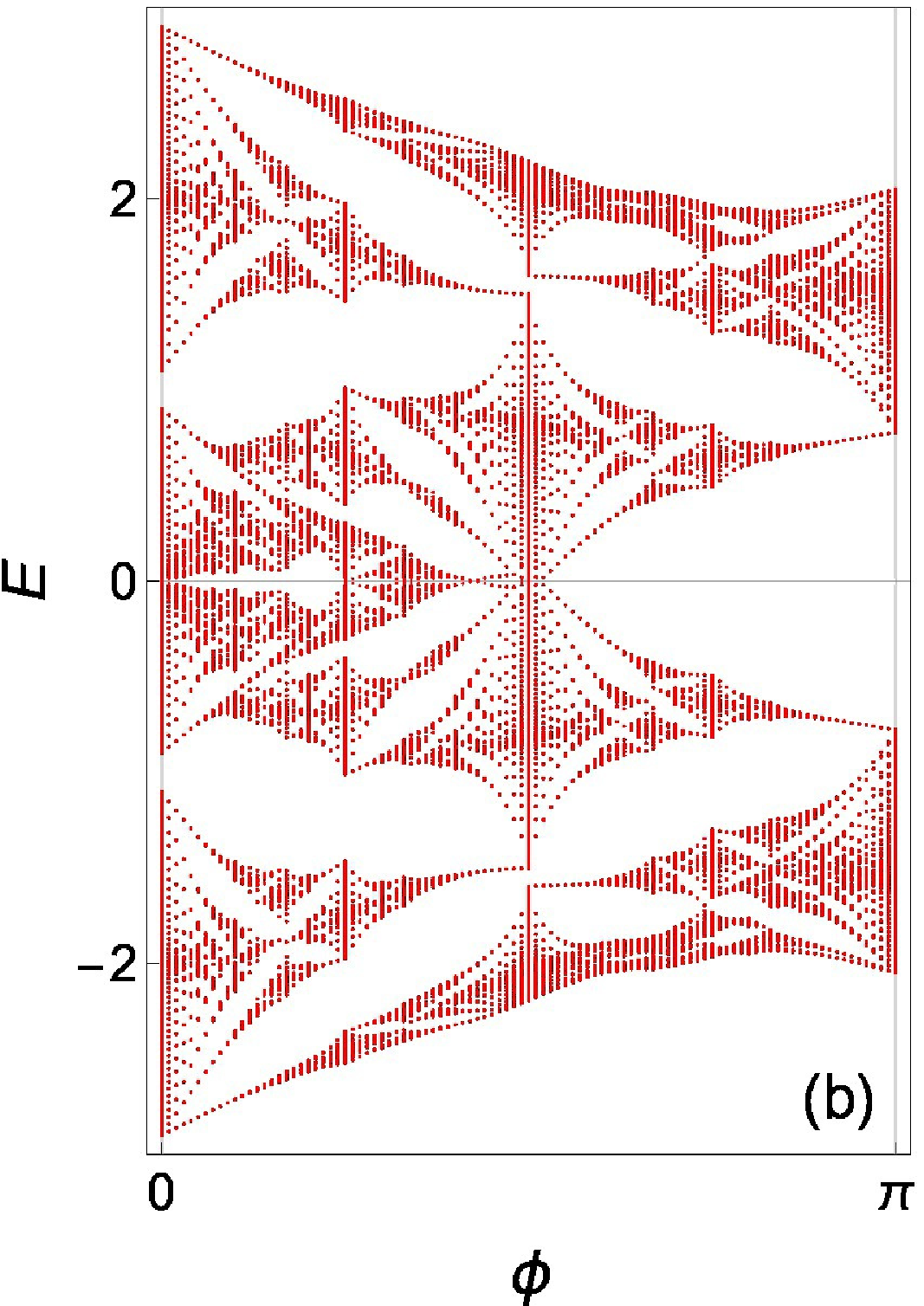}
\end{tabular}
\caption{
Hofstadter butterfly spectra for (a) $\gamma=0.35$ and (b) $\gamma=0.45$.
The spectra in $\pi<\phi<2\pi$ are symmetric with respect to $\phi=\pi$.
}
\label{f:hof}
\end{center}
\end{figure}
(2) As $\gamma$ becomes larger, the gap becomes smaller, and around $\gamma\sim0.4$, a gap-closing
occurs at $\phi=\pi/2$. 
Even after the gap-closing, one can observe a smaller but finite gap
surviving in the weak field regime $0<\phi<\pi/2$, as can be seen in Fig. \ref{f:hof}.
On one hand, with $\phi$ fixed, the states in this gap can be continuously deformed into those 
with smaller $\gamma$, and eventually reach those in the leftmost panel of Fig. \ref{f:series_Hof} without any gap-closings.
This implies that they are HOTI.
%
On the other hand, with $\gamma$ fixed,
they can no longer be deformed into those around $\phi=\pi$ due to a gap-closing around $\phi\sim\pi/2$.
Therefore, it is desirable to determine their topological properties directly.
(3) As $\gamma$ becomes much larger, the gap around $\phi=\pi$ shrinks and 
eventually vanishes at $\gamma=1\,(=\lambda)$.
Then, the HOTI phase disappears from the butterfly. 

\begin{figure}[htb]
\begin{center}
\begin{tabular}{cc}
\includegraphics[width=.49\linewidth]{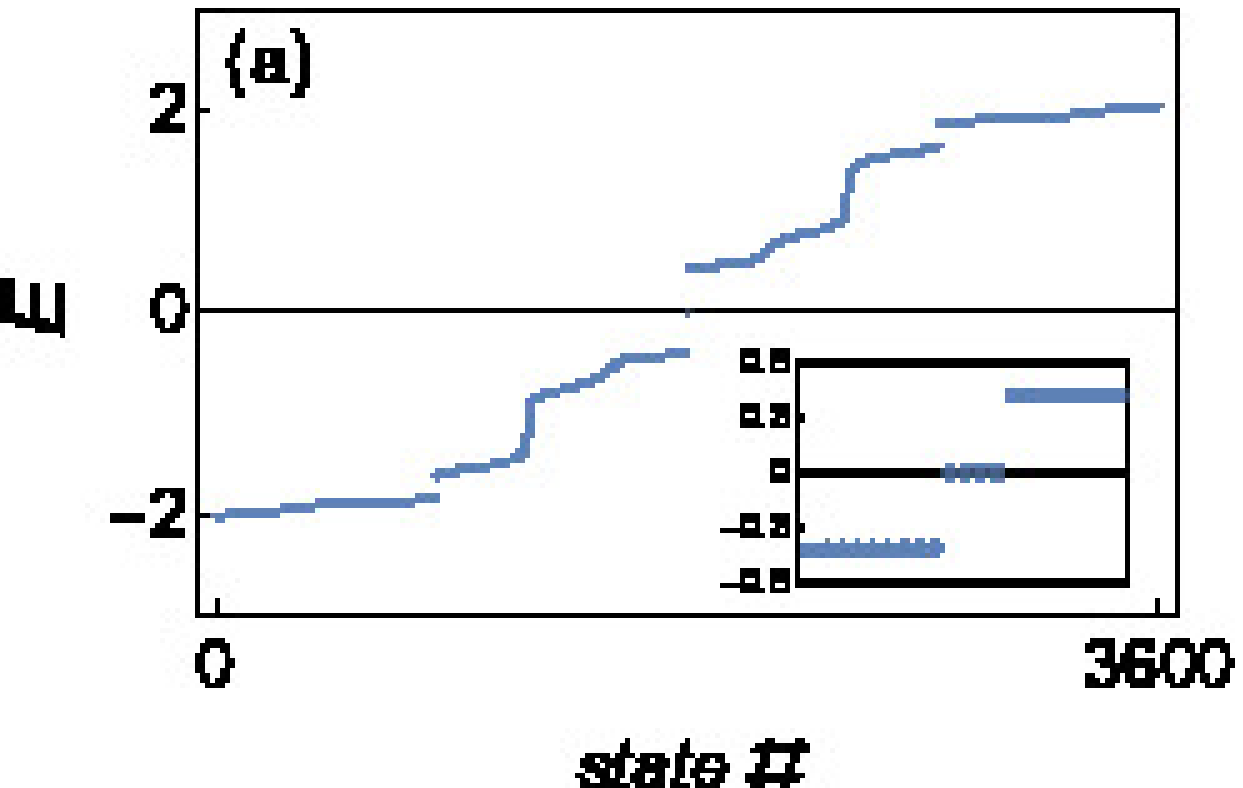}
&
\includegraphics[width=.49\linewidth]{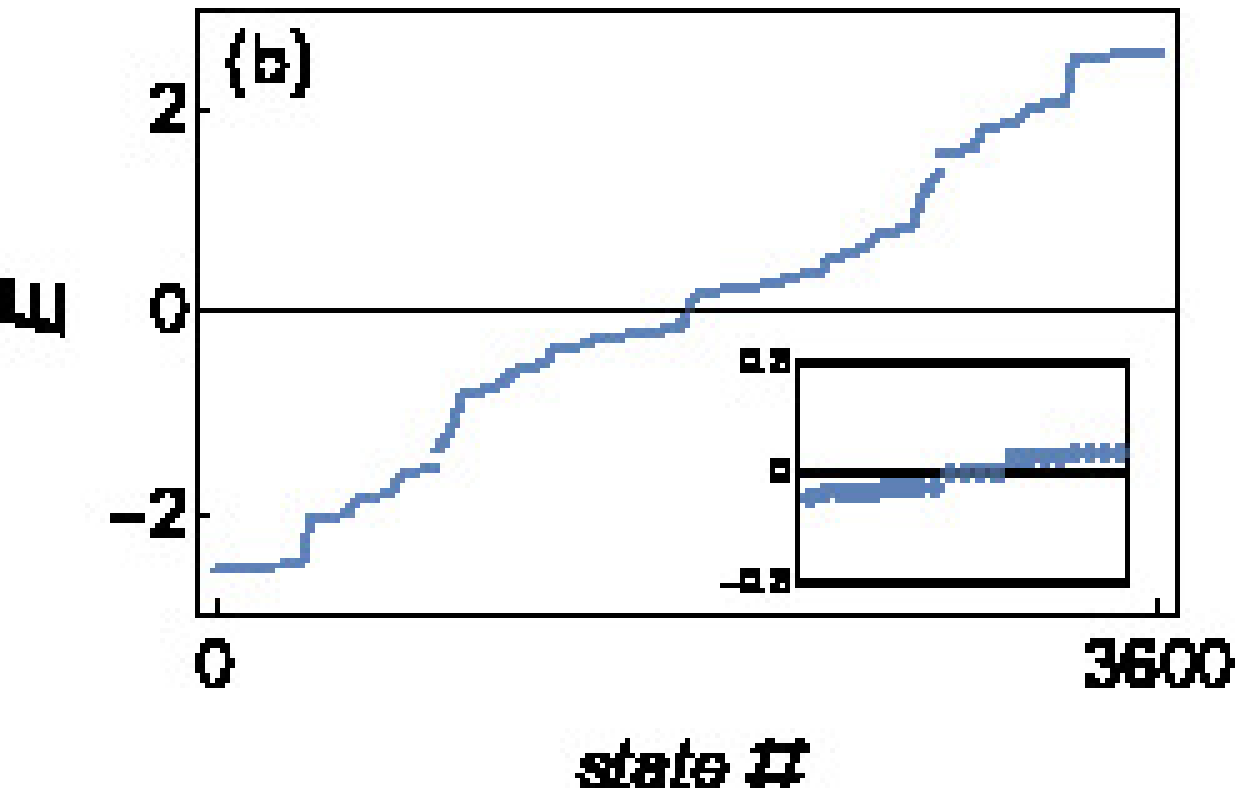}
\\
\includegraphics[width=.49\linewidth]{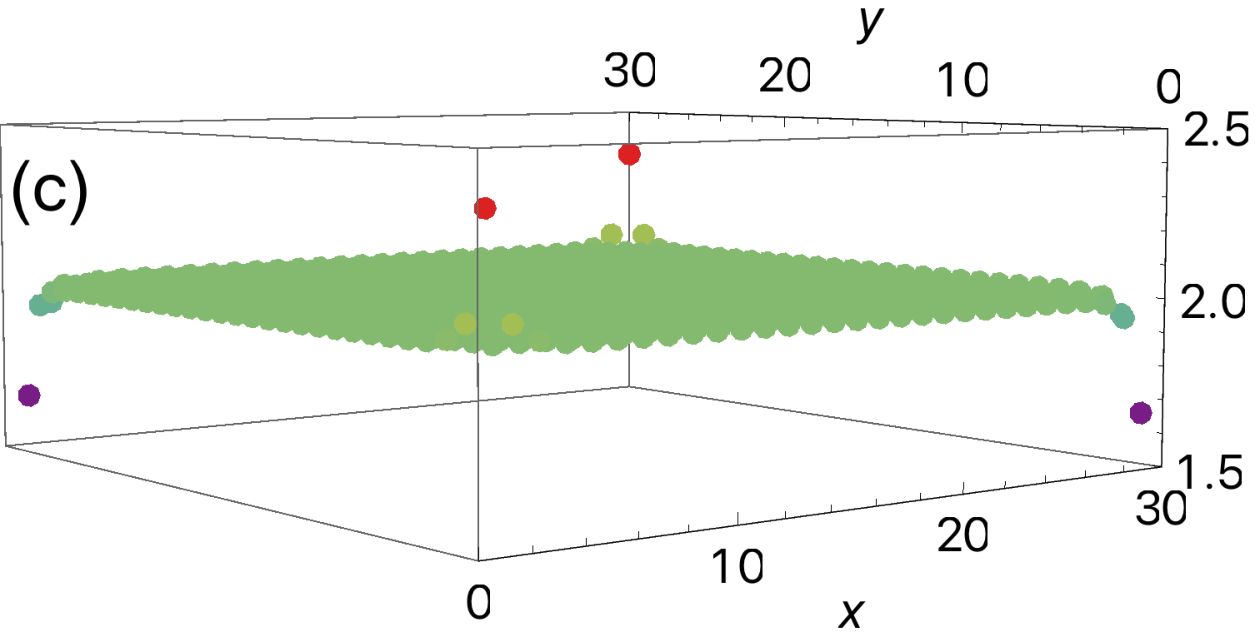}
&
\includegraphics[width=.49\linewidth]{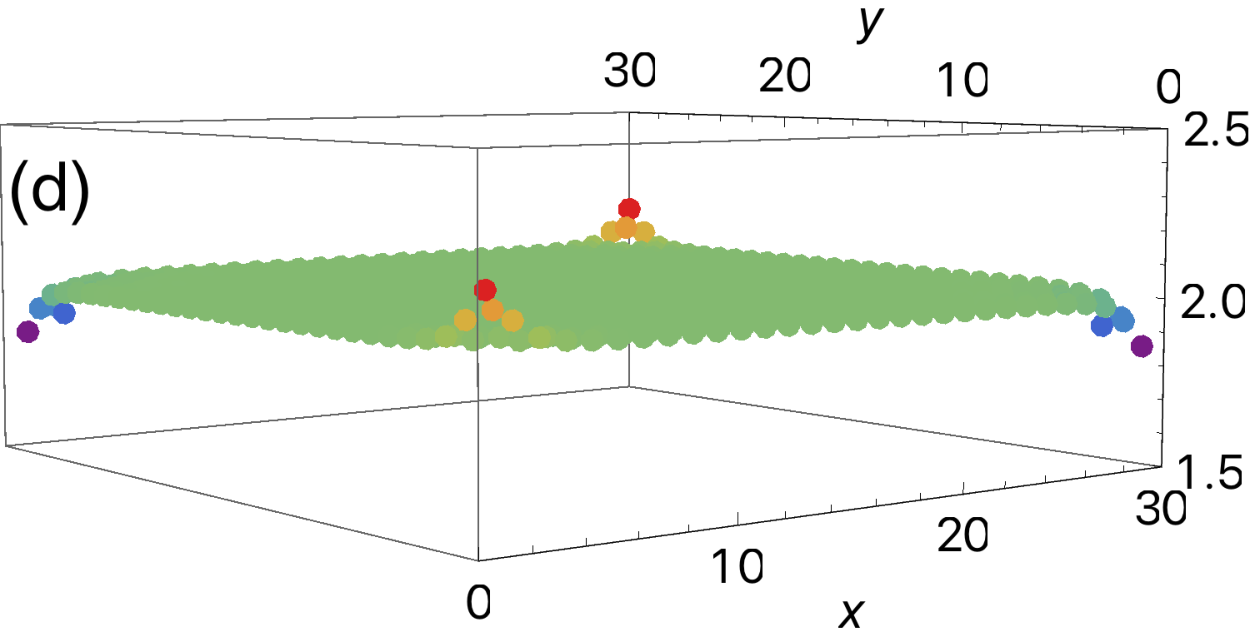}
\end{tabular}
\caption{
(a, b) The energies of the model with full open boundary conditions in the case of $\gamma=0.45$.
Insets show degenerate four zero energy states.
(c, d) Occupied charge per SSH unit cell.
Left (a, c) and right (b, d) are under  flux $\phi=2\pi/3$ and $\phi=2\pi/10$, respectively. 
The system size is $60\times60$ sites ($30\times30$ SSH unit cells).
}
\label{f:corner}
\end{center}
\end{figure}

\subsection{Corner states}\label{s:cor}

In what follows, we restrict our discussions to the model with $\gamma=0.45$ in Fig. \ref{f:hof}(b), 
and calculate the corner charges at two values of magnetic flux $\phi$.
In Fig. \ref{f:corner}, we show the spectrum of the model with full open boundary conditions.
Figure \ref{f:corner} (a) is the case with flux $\phi=2\pi/3$, which may be adiabatically connected to the BBH model.
One observes degenerate four zero energy states in the bulk gap 
whose wave functions are localized at four corners, as seen in Fig. \ref{f:corner}(c).
These give indeed corner charge $\sim\pm 0.494$ (charge deviation from 2 within $3\times3$ unit cells around each corner).
This suggests that the gapped ground state belongs to topological quadrupole phase with 
corner charges $\pm1/2$.

Even in the isolated gap in the weak field regime,  one can also observe
degenerate  zero energy states in a small energy gap in Fig. \ref{f:corner}(b). 
Although the peaks and valleys of the occupied charges at corners do not look very sharp in Fig. \ref{f:corner}(d), 
the corner charge can be estimated as $\pm 0.435$ and 
$\pm 0.492$, respectively, within  $3\times3$ and $5\times5$ unit cells around each corner.
Therefore, the distribution of the corner charge is rather broad, but its total amount
would be $\pm 1/2$. 
Thus, the question is whether the states in the weak field regime belong to the HOTI phase.
To address the question, we apply the entanglement techniques to this system developed in Refs. \cite{Fukui:2014qv,Fukui:2015fk}
and 
applied to the HOTI phase of the BBH model \cite{Fukui:2018aa}. 
Also important is the stability of the zero energy states among an extremely small gap, which will be 
discussed in Sec. \ref{s:ex}.

\section{Entanglement polarization}\label{s:ep}

In this section, we introduce the eP which can be  topological invariants characterizing the HOTI. 
We discuss first the eP for the bulk in Sec. \ref{s:ep_bulk}, and next the eP for the edge states 
in Sec. \ref{s:ep_edge}, which is useful to lift the
degeneracy of the edge states and characterize each of them separately.

\subsection{Entanglement polarization for the bulk}\label{s:ep_bulk}

We  divide the total system into a subspace $A$ and its complement $\bar{A}$, and derive the entanglement 
Hamiltonian (eH) $H^{A}$ and $H^{\bar{A}}$ as follows.
Let $|G\rangle$ be the half-filled ground state of the model with a flux $\phi$.
Then, by tracing out $\bar A$ in the density matrix $\rho=|G\rangle\langle G|$, 
we obtain eH, ${H}^{A}$, as
$\tr_{\bar{A}}\,\rho\propto e^{-H^{A}}$.
Since for noninteracting systems, the eH thus defined also reduce to 
noninteracting  Hamiltonians \cite{Peschel:2003uq}, one can define the Berry connections 
associated with the eigenfunctions of the eH, which are denoted by $A^A_\mu(k_x,k_y)$ with $\mu=x,y$.
The integration of $A^A_x(k_x,k_y)$ and $A_y^A(k_x,k_y)$, respectively, over $k_x$ and $k_y$
defines the eP, $p^A_x$ and $p^A_y$ \cite{Fukui:2018aa}.
For details, see the Appendix.

To characterize the HOTI, we introduce two kinds of partitions:
One is $A=\{1,2,\ldots,q\}\equiv\,\downarrow$ sites,
and the other is $A=\{1,3,\ldots,q-1,q+1,q+3,\ldots,2q-1\}\equiv{L(\rm eft)}$ sites in the magnetic unit cell in  Fig. \ref{f:lat}. 
Here, left means the left sites in the SSH unit cell.
Their complements are denoted as $\uparrow$ and $R$(ight), respectively.
In what follows, we often use $\sigma=\,\downarrow$ or $\uparrow$ and $\tau={L}$ or $R$, and $-\sigma$ and $-\tau$ stand for 
the complement of $\sigma$ and $\tau$, respectively. 
As shown in the Appendix, the symmetry properties Eq. (\ref{Sym}) force such eP quantized as $0$ or $1/2$.
Thus, the set of bulk eP, $(p^\sigma_x,p^\tau_y)$, can be topological invariants 
characterizing the HOTI.

Let us calculate eP for the two cases in Fig. \ref{f:corner}.
First of all, we mention that the eS under 
$\phi=2\pi/3$ and $\phi=2\pi/10$ are indeed gapped, although the gap under $\phi=2\pi/10$ is 
rather small. 
Therefore, it is possible to compute eP for occupied state $(\xi>1/2)$.
Using the link variable technique for the Berry connections \cite{FHS05}, 
we have $(p^\sigma_x,p^\tau_y)=(1/2,1/2)$
in both cases $\phi=2\pi/3$ and $\phi=2\pi/10$.
The nontrivial $p^\sigma_x=1/2$ implies that the 1D single chain toward the $x$ direction specified by $\sigma$,
if disentangled from another chain $-\sigma$, is topologically equivalent to the
SSH chain with edge states. 
Thus, each chain $\sigma$ or $-\sigma$ has potentially edge states at zero energy.
However, the edge states are lifted toward nonzero energies as a result of the coupling between the chains 
$\pm\sigma$.
These form gapped edge states, still localized near the edges along the $y$ direction. 
What $p^\tau_y=1/2$ means is likewise. 

\begin{figure}[htb]
\begin{center}
\begin{tabular}{cc}
\includegraphics[width=.49\linewidth]{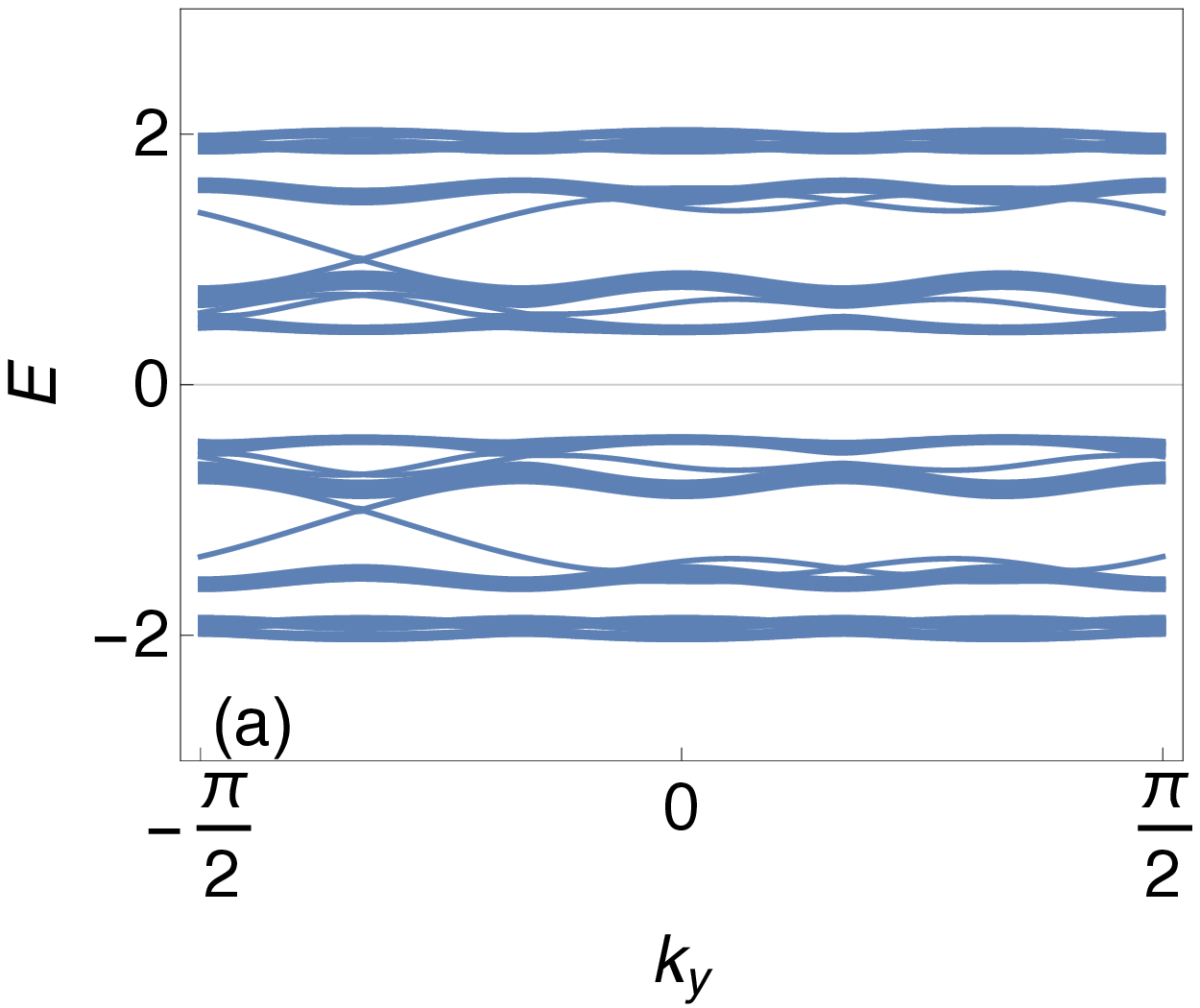}
&
\includegraphics[width=.49\linewidth]{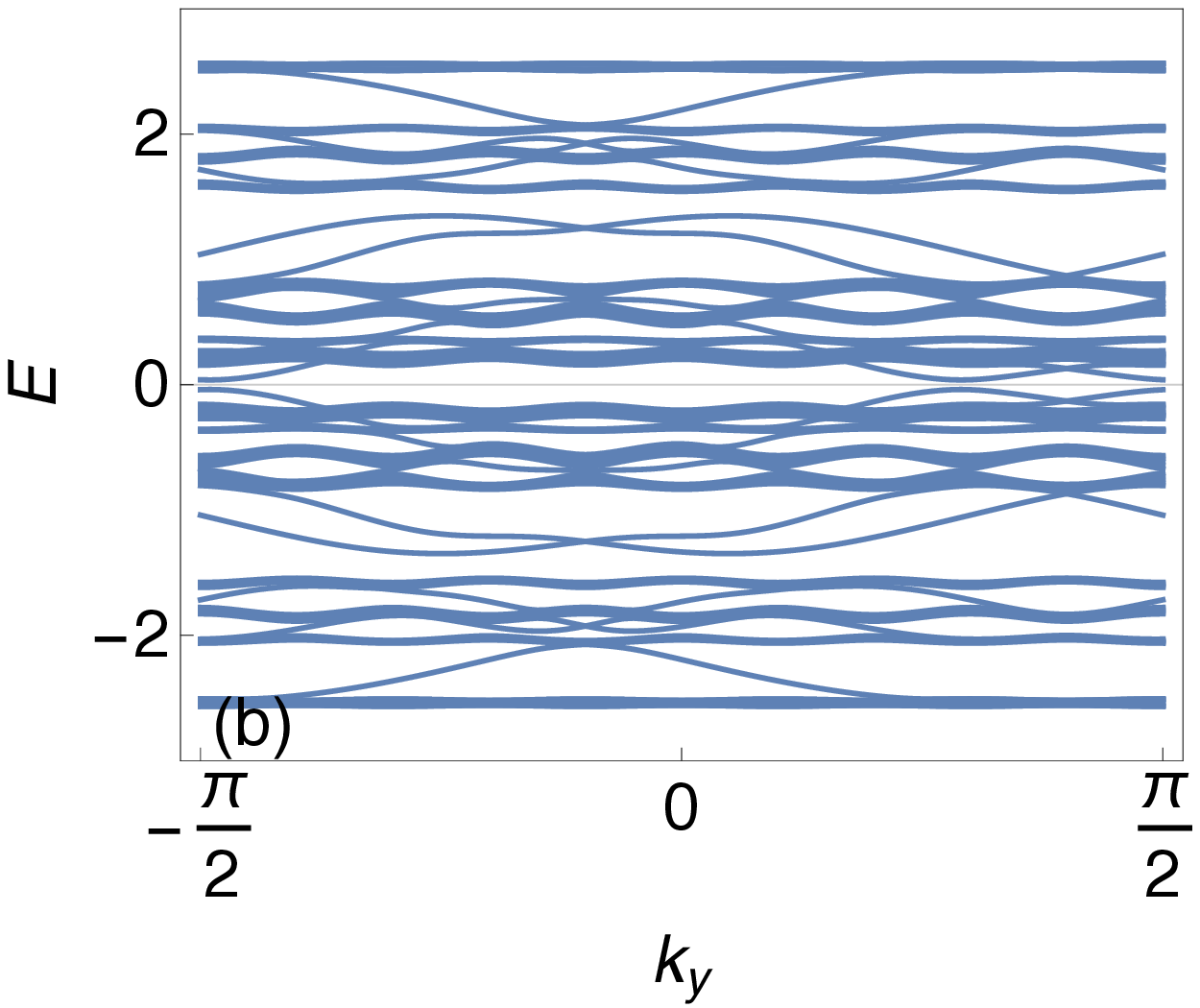}
\end{tabular}
\caption{
Spectra of the model with open (periodic) boundary condition toward the $x$ ($y$) direction, 
in the case of $\gamma=0.45$. (a, b) are under
flux $\phi=2\pi/3$ and $\phi=2\pi/10$, respectively. 
}
\label{f:edge_x}
\end{center}
\end{figure}

\subsection{Entanglement polarization for the edge states}\label{s:ep_edge}

So far we derived the eP for the bulk ground states. 
The two bulk ground states have indeed nontrivial eP, implying the existence of 
gapped edge states, if the open boundary condition is imposed in one direction.
This is one of the characteristic properties of the HOTI. 
Therefore, we switch our attention to the discussion of the edge states.

\subsubsection{Edge states}

In Fig. \ref{f:edge_x}, we show the spectra of the model with open boundary condition in the $x$ direction.
Since each Landau level  has a nontrivial Chern number, one can observe various 
edge states in between the Landau levels at nonzero energy, but {\it no edge states across the zero energy. }
Therefore, gapped edge states associated with the SSH zero energy states, even if they exist, are embedded
somewhere in the spectra, although it is hard to distinguish them from others. 
Even if these states are identified, they are spectrally degenerate, 
although spatially separated at the left and right ends,  as in the case of the BBH model.

\subsubsection{Entanglement edge state polarization}

It should be noted that the entanglement technique, applied to the system with boundaries,  
enables us to carry out such an identification of the single edge state
and to compute its eP, which may be referred to as entanglement edge state polarization (eESP).
To this end, 
let us construct the projection operator $P_{\rm G}$ in Eq. (\ref{ProOpeG}) 
using the wave functions of the ground states 
with open boundary condition in the $x$ direction, i.e., those of Fig. \ref{f:edge_x}.
Let us introduce similar partitions $A=\sigma$, or $A=\tau$, extending the magnetic unit cell into 
whole finite chains in the $x$ direction.
Then, we obtain $P_{\rm G}^A(k_y)=P^AP_{\rm G}(k_y)P^A$, 
from which we compute the eS and eP including the edge states.

\begin{figure}[htb]
\begin{center}
\begin{tabular}{cc}
\includegraphics[width=.49\linewidth]{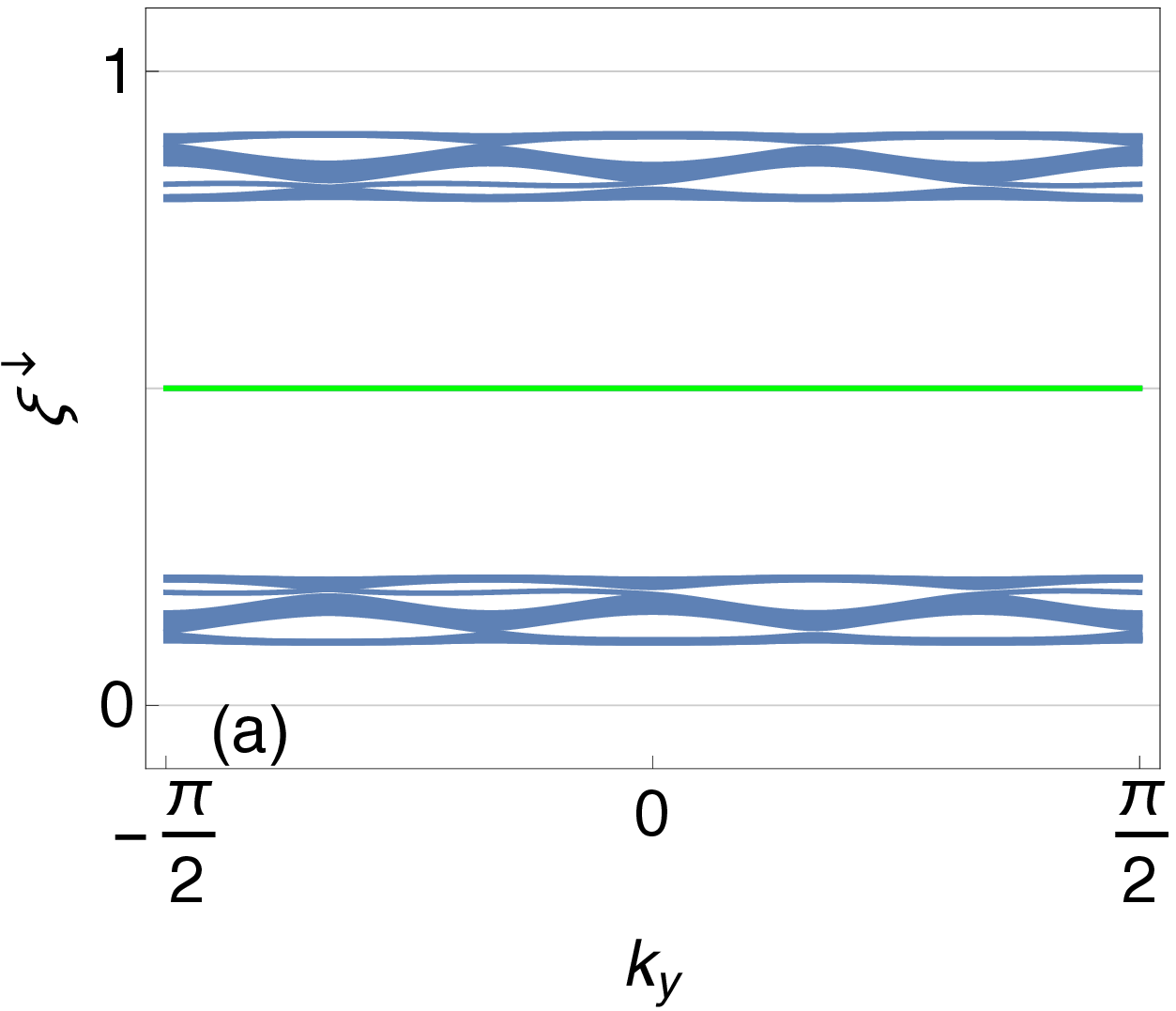}
&
\includegraphics[width=.49\linewidth]{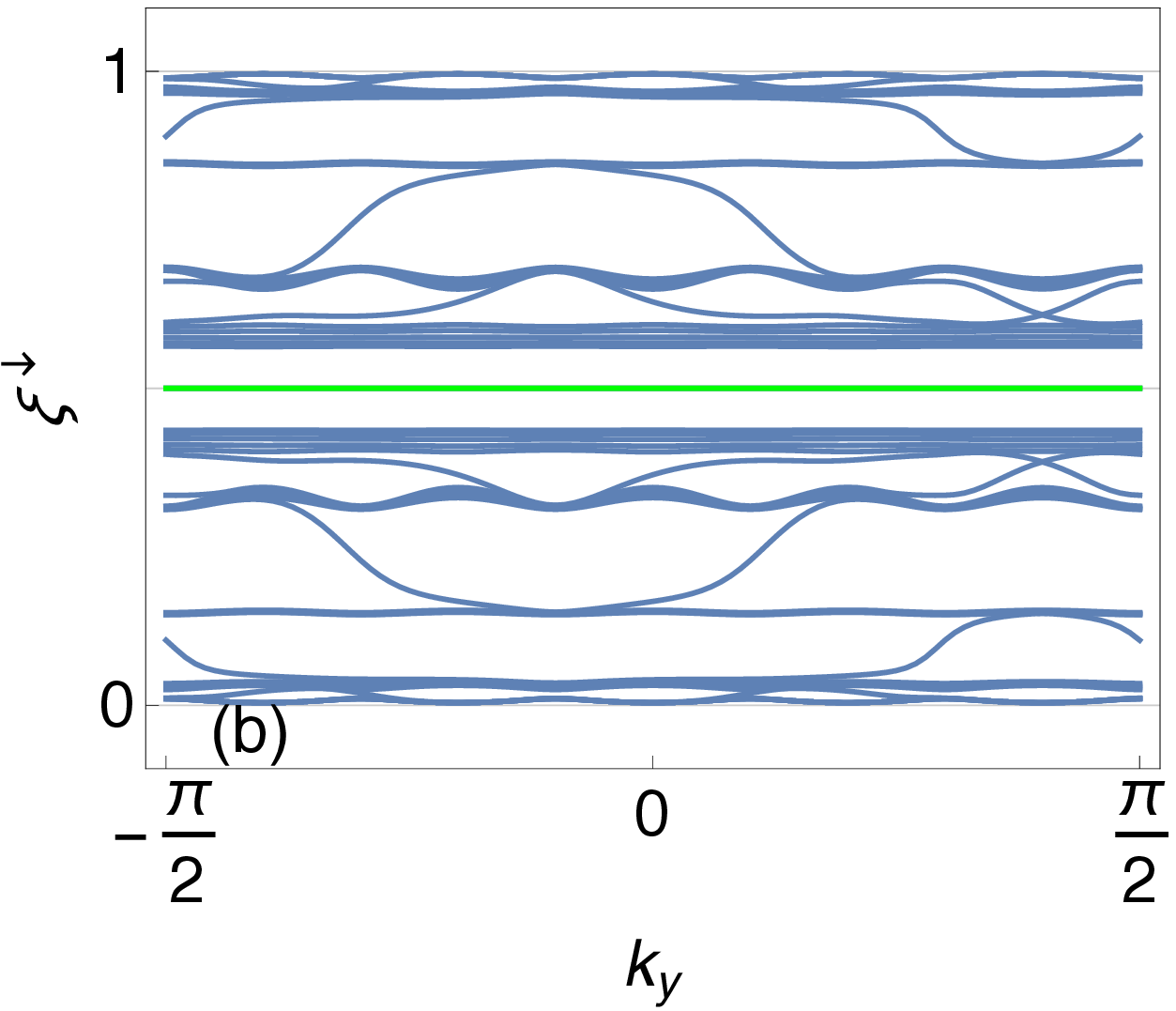}
\\
\includegraphics[width=.49\linewidth]{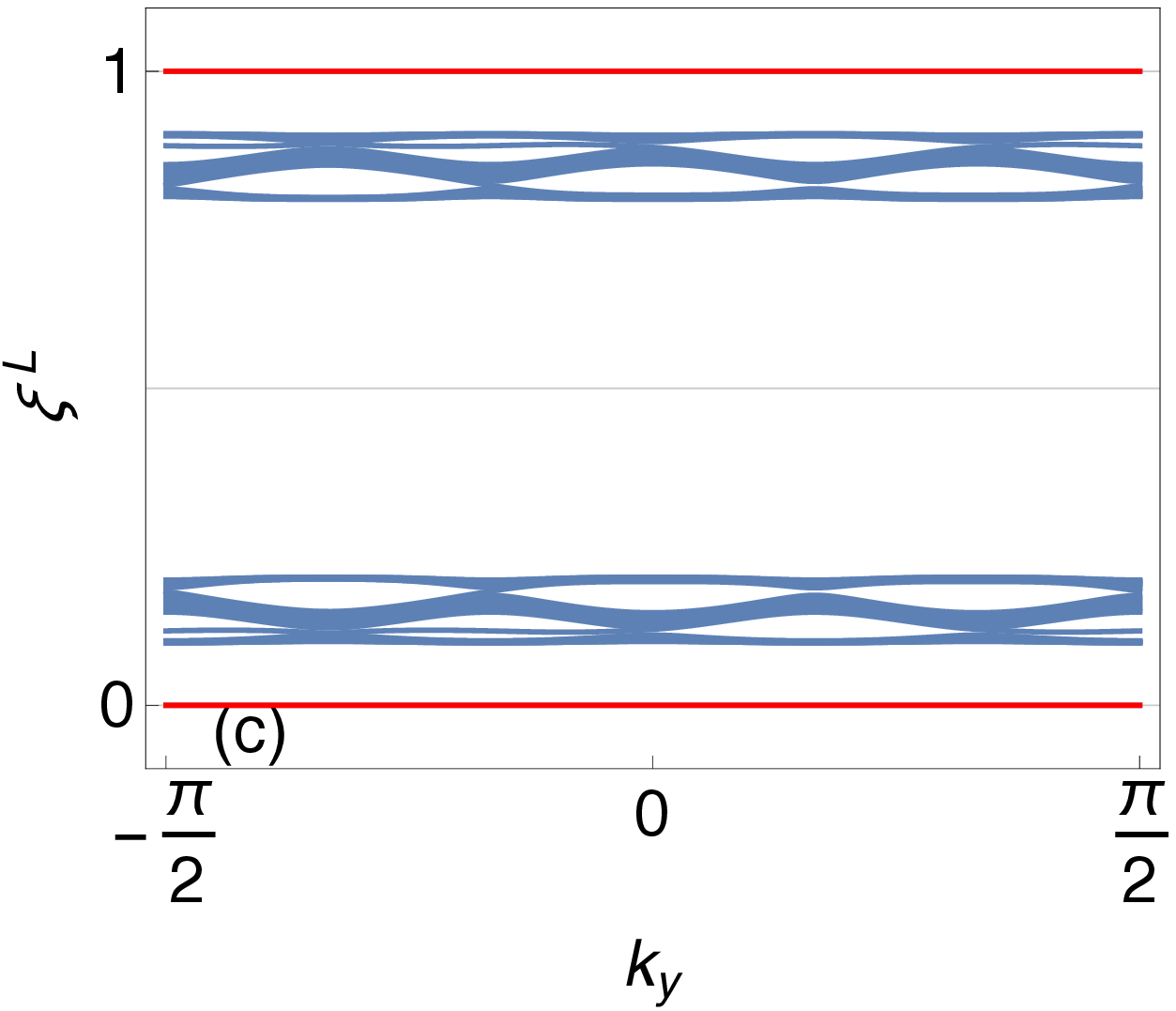}
&
\includegraphics[width=.49\linewidth]{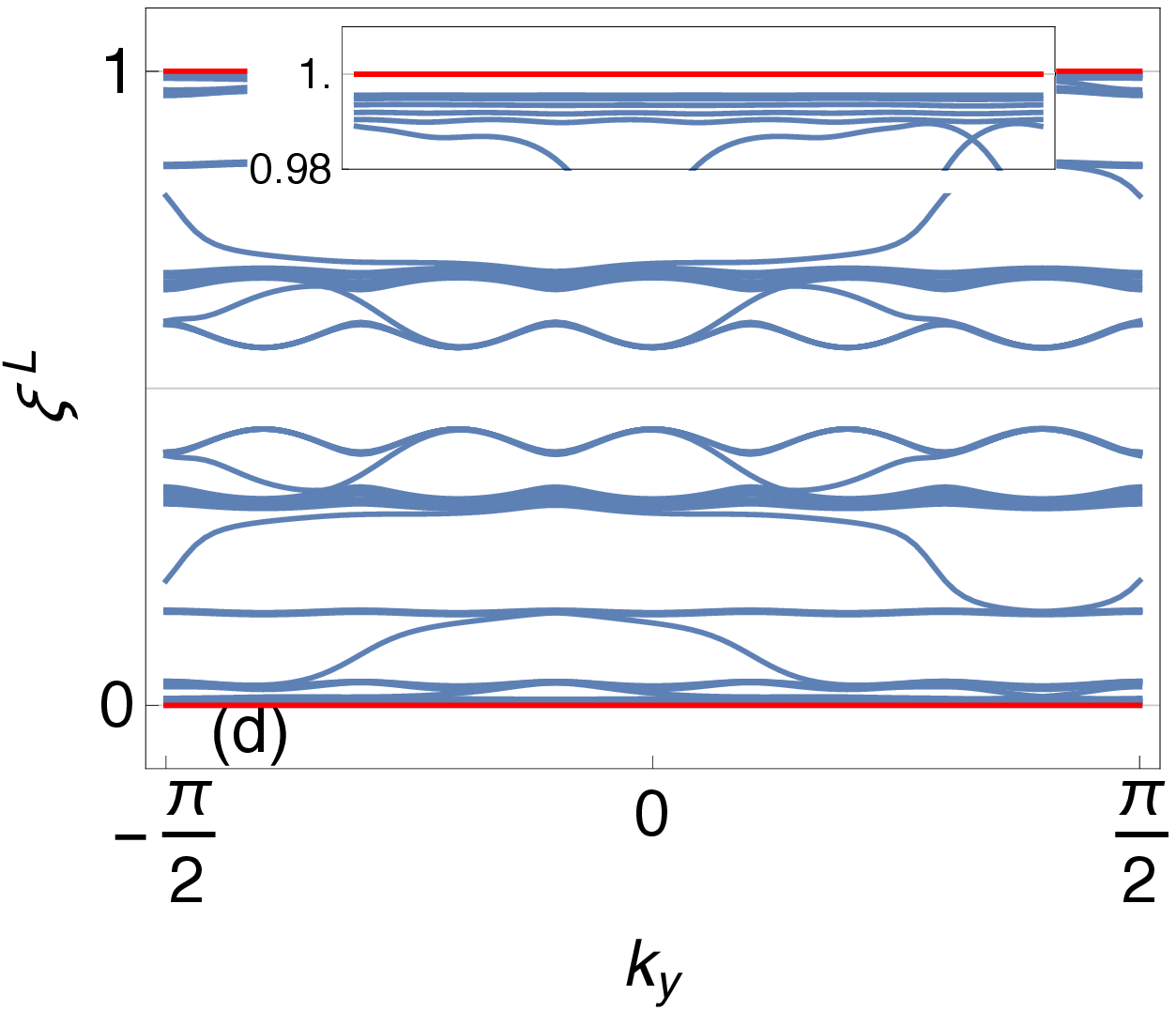}
\end{tabular}
\caption{
The entanglement spectra for Fig. \ref{f:edge_x}. Left (a, c) and right (b, d) are under
flux $\phi=2\pi/3$ and $\phi=2\pi/10$, respectively. 
The inset shows the spectrum near $\xi=1$.
}
\label{f:ent_edge}
\end{center}
\end{figure}

In Figs. \ref{f:ent_edge}(a) and \ref{f:ent_edge}(b), we show the eS, $\xi^{\downarrow}(k_y)$. 
One can clearly observe (doubly-degenerate) zero energy states indicated by the green lines.
Thus, we can reproduce the zero energy edge states in the topological SSH phase. 
To check this argument, let us introduce anisotropy of the hopping parameters.
First, consider the system with $(\gamma_x,\lambda_x)=(0.45,1)$ and $(\gamma_y,\lambda_y)=(1,0.45)$ which has the bulk eP,
$(p^\sigma_x,p^\tau_y)=(1/2,0)$. 
This case has the same spectra
$\xi^\downarrow(k_y)$ as in Figs. \ref{f:ent_edge}(a) and \ref{f:ent_edge}(b) with zero energy states.
\begin{figure}[htb]
\begin{center}
\begin{tabular}{cc}
\includegraphics[width=.49\linewidth]{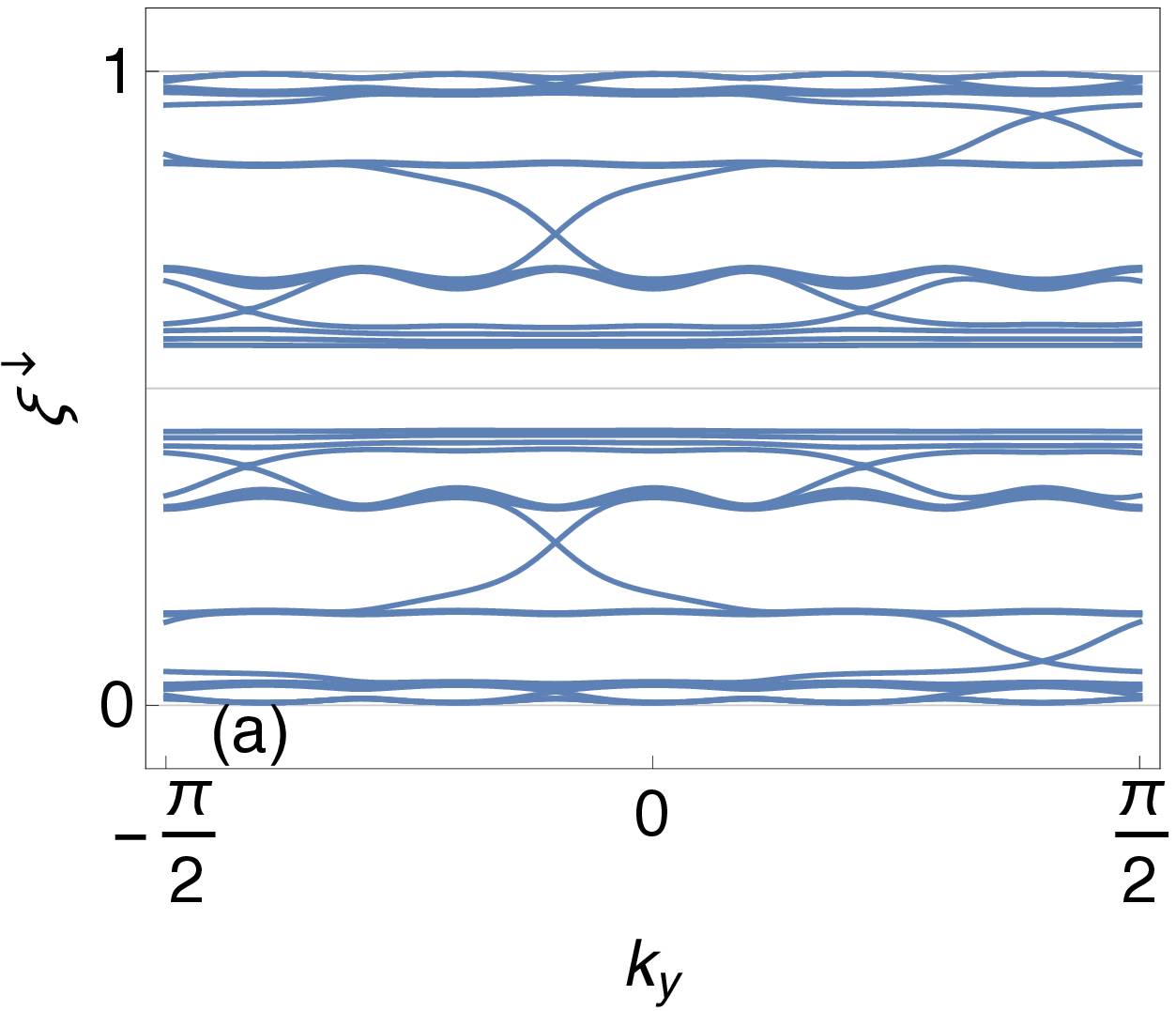}
&
\includegraphics[width=.49\linewidth]{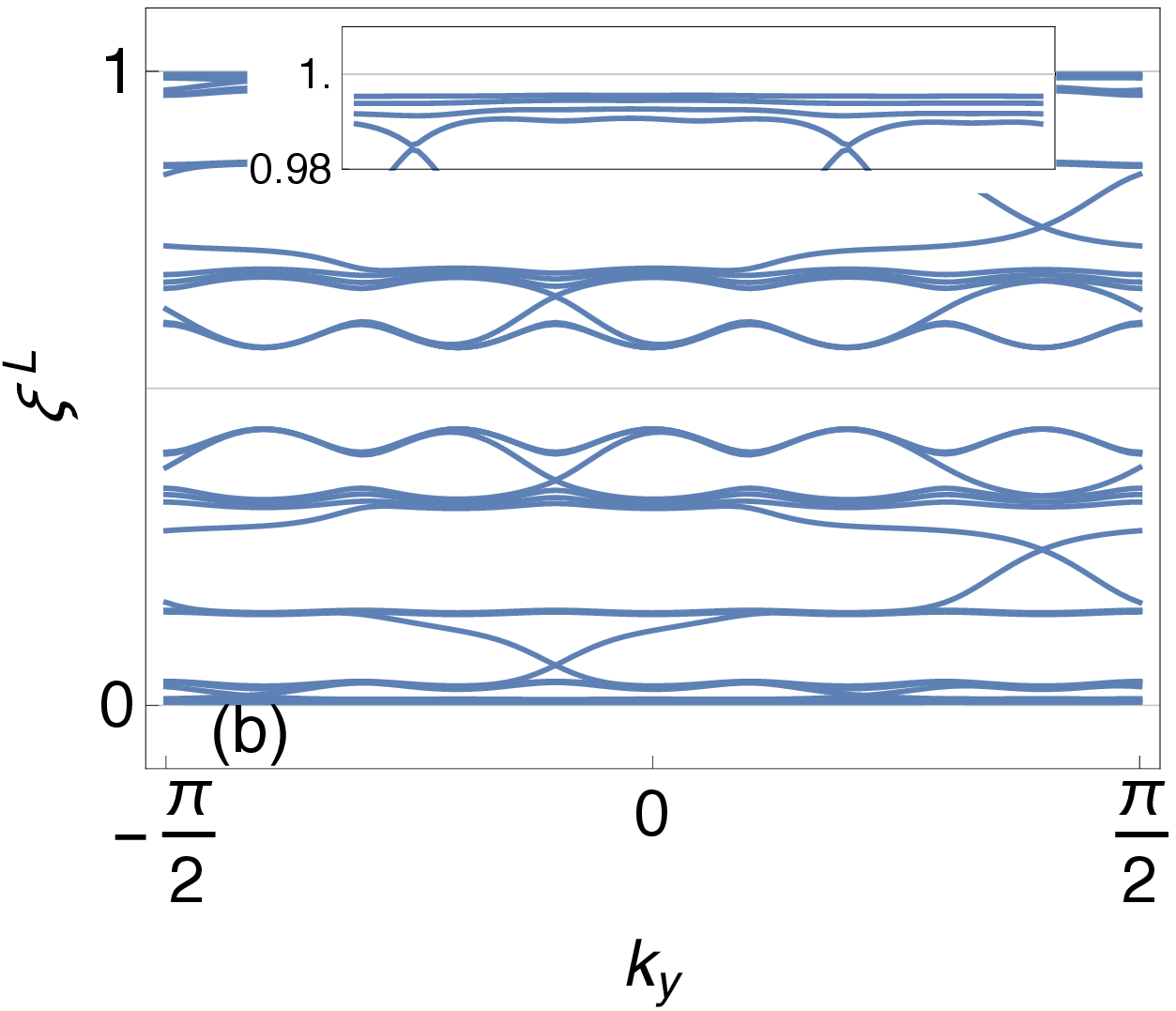}
\end{tabular}
\caption{
eS with anisotropic hopping parameters $(\gamma_x,\lambda_x)=(1,0.45)$ and $(\gamma_y,\lambda_y)=(0.45,1)$ under $\phi=2\pi/10$. 
}
\label{f:anisotropic}
\end{center}
\end{figure}
Second, consider the system with $(\gamma_x,\lambda_x)=(1,0.45)$ and $(\gamma_y,\lambda_y)=(0.45,1)$
which has the bulk eP, $(p^\sigma_x,p^\tau_y)=(0,1/2)$. This case shows similar spectra but with no zero energy states,
as in Fig. \ref{f:anisotropic}(a), where only the case with $\phi=2\pi/10$ is shown.
Thus, disentanglement between two chains $\sigma=\,\uparrow, \downarrow$ enables to reveal the gapped edge states 
associated with each 1D chain as the zero energy edge states.

To reveal the property of these gapped edge states,
let us next consider the partition $\tau$,
which lifts the degeneracy of gapped edge states at the left and right ends as follows:
For example, $\tau=L$
includes only the left end. Therefore, the edge states localized at the left and right ends are, respectively,  almost 
occupied and unoccupied in the partition $\tau=L$.
Thus, we can spectrally separate the edge states at the left and right ends.
In Figs. \ref{f:ent_edge}(c) and \ref{f:ent_edge}(d), we show the eS, $\xi^{L}(k_y)$, for the partition $\tau={L}$. 
One can observe $\xi=1$ and $\xi=0$ states indicated by red lines.
For these states, numerical calculations of the eP show that 
we obtain $1/2$ eESP both for $\xi=1$ and $\xi=0$ states, implying that these are 1D SSH topological states propagating 
toward the $y$ direction along the edges perpendicular to the $x$ direction.
Therefore, if the open boundary condition is further imposed in the $y$ direction, zero energy edge states
appear. These are nothing but the corner states.

To check this argument, let us again introduce anisotropic hopping parameters.
The system with $(\gamma_x,\lambda_x)=(0.45,1)$ and $(\gamma_y,\lambda_y)=(1,0.45)$ has the same spectrum 
$\xi^{L}(k_y)$ with $\xi=1$ and $\xi=0$ states. However, their eESP are 0, implying that these gapped edge states
are trivial dimerized states. Therefore, 
even if the open boundary condition is further imposed in the $y$ direction, no edge states appear. 
The system with $(\gamma_x,\lambda_x)=(1,0.45)$ and $(\gamma_y,\lambda_y)=(0.45,1)$ shows no 
$\xi=1$ and $\xi=0$ states, as shown in Fig. \ref{f:anisotropic}(b).

\section{Experimental feasibility}\label{s:ex}
In this section, we briefly discuss the experimental feasibility of the HOTI.
As we showed in Sec. \ref{s:model}, a wide gap is open for $0<\phi\le\pi$ at half-filling, if $\gamma$ is small.
However,  this gap converges to 0 for $\phi\rightarrow0$, as can be seen in Fig. \ref{f:series_Hof}, implying that
the simple SSH model with $\phi=0$ can be considered as a critical point of the HOTI phase.
Near this critical point, the HOTI  seem to be unstable against small perturbations and/or disorder.  

\begin{figure}[htb]
\begin{center}
\begin{tabular}{cc}
\includegraphics[width=.47\linewidth]{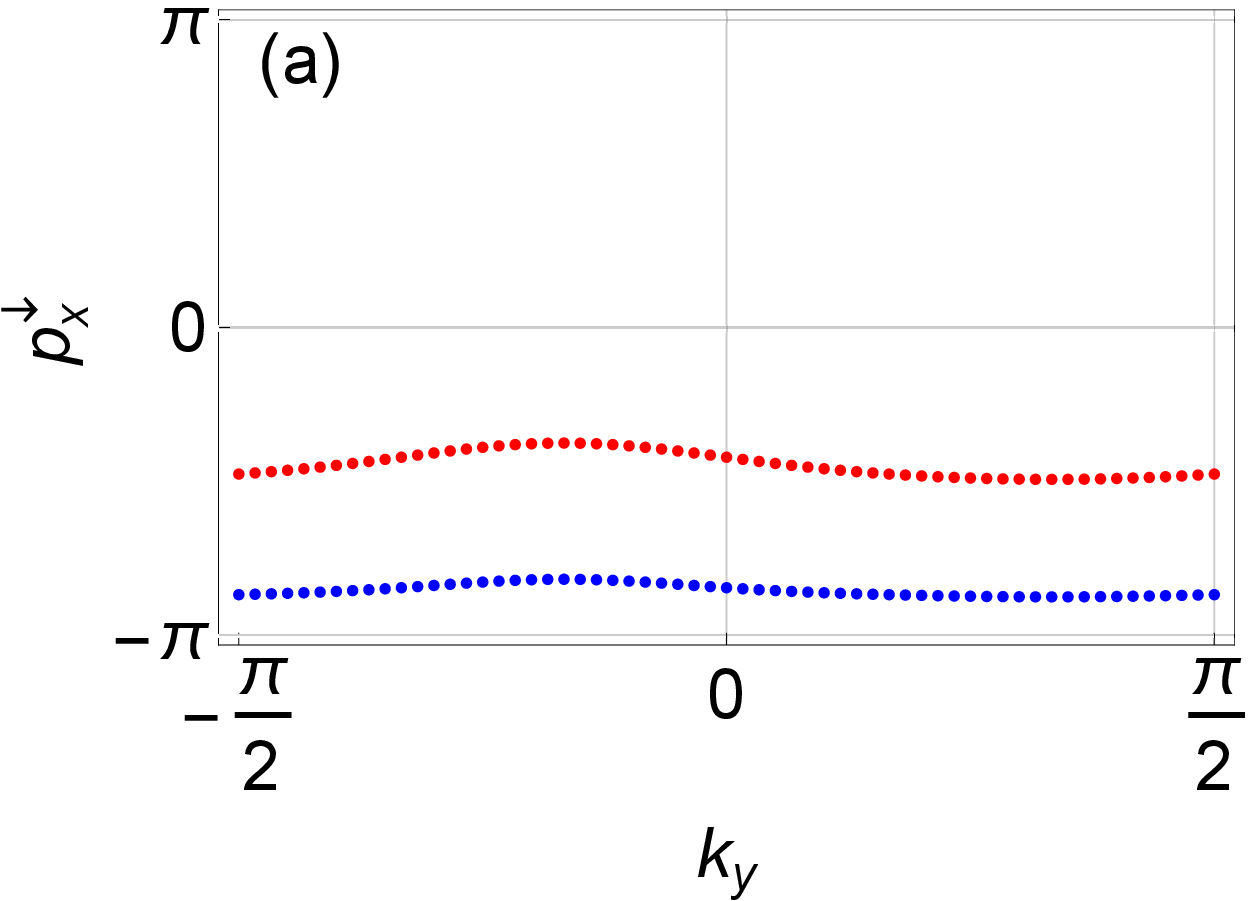}
&
\includegraphics[width=.52\linewidth]{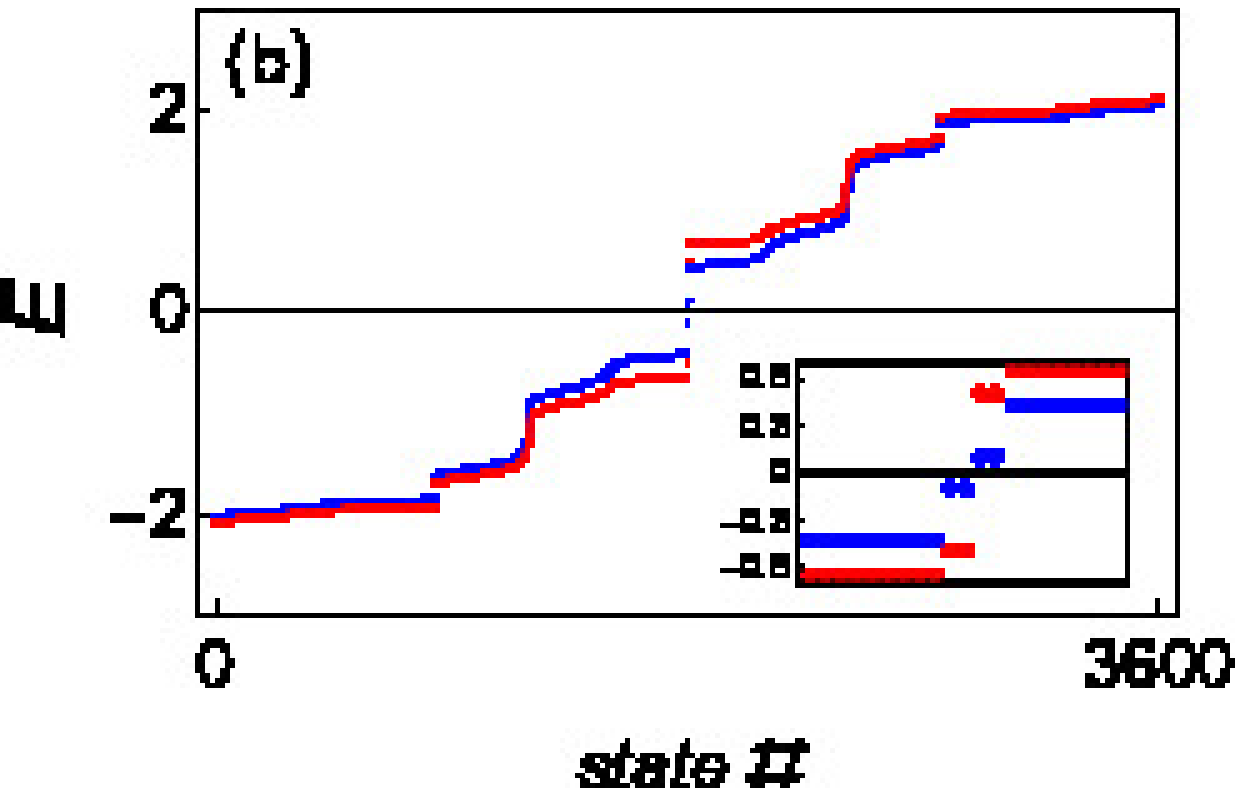}
\\
\includegraphics[width=.49\linewidth]{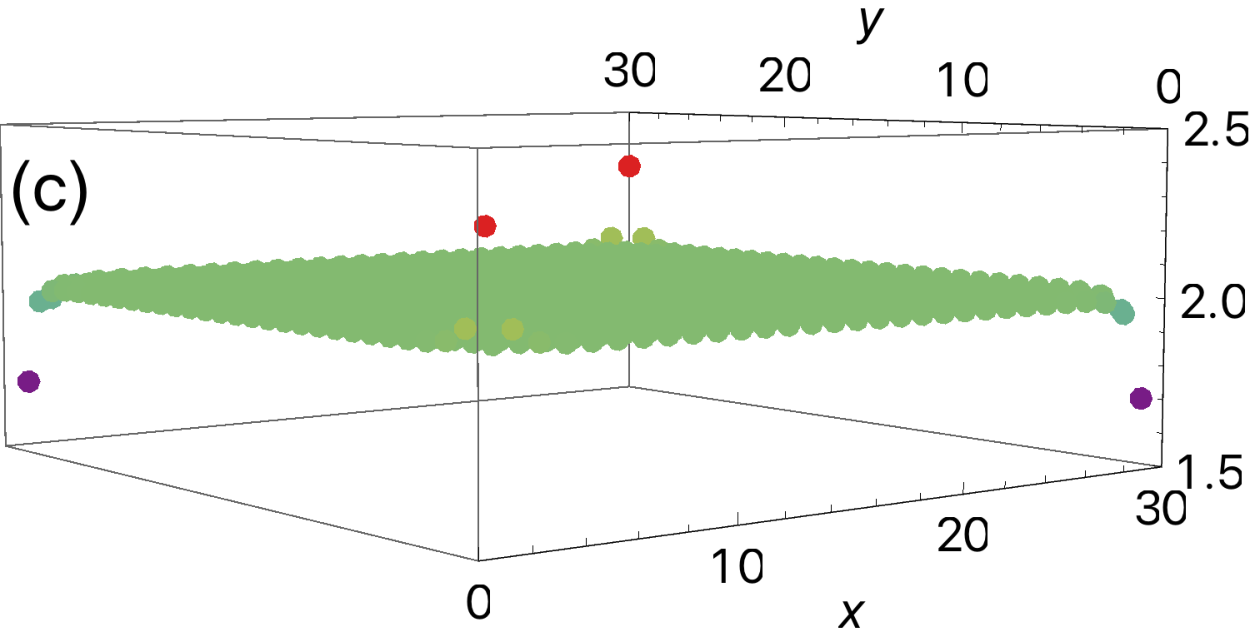}
&
\includegraphics[width=.49\linewidth]{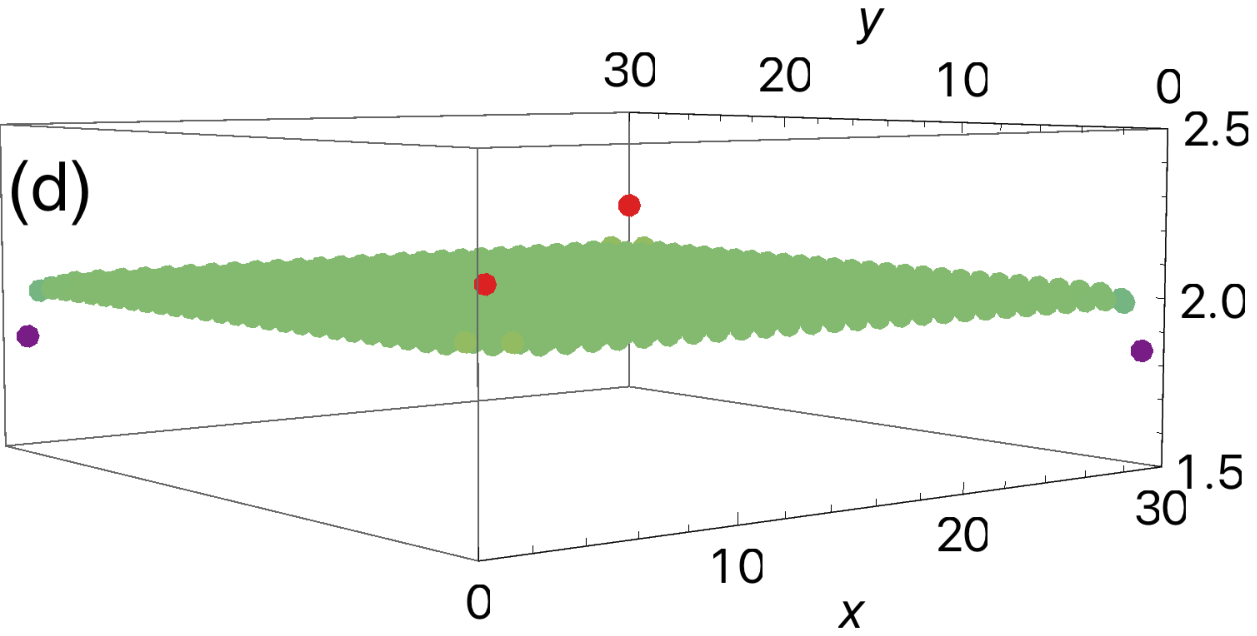}
\end{tabular}
\caption{
(a) eP of the bulk, $p_x^\downarrow$, and  (b) energies with full open boundary conditions, 
corresponding to Figs. \ref{f:corner}(a) and \ref{f:corner}(c) ($\gamma=0.45$ $\phi=2\pi/3$).
The staggered potential is included, blue $\delta=0.1$ and red $\delta=0.5$.
(c, d) show the occupied charges per SSH unit cell with $\delta=0.1$ and $\delta=0.5$, respectively.
}
\label{f:bre1}
\end{center}
\end{figure}

However, as we will discuss below, the HOTI phase is robust against the staggered potential which opens a large gap,
\begin{alignat}1
H_{\rm st}=\delta\sum_j (-1)^{j_x+j_y} c_j^\dagger c_j,
\label{Sta}
\end{alignat}
and therefore, there may still be the possibility of experimental observations of the corner states in a weak field regime.
Since the staggered potential breaks reflection symmetries (\ref{OriTraLaw}) and (\ref{Sym}), 
the eP are no longer quantized, and hence, the symmetry-protected corner states would vanish continuously as $\delta$ increases.
Nevertheless, if the symmetry-protected states without staggered potentials, which is referred to as {\it the mother states}, 
are in the HOTI phase, 
the corner states may be robust enough; we could observe their signature in experiments, as will be demonstrated below. 
Furthermore, the staggered potential yields  a large gap, which may stabilize the corner states against disorder.

Let us start illustrating such an indirect observation of the corner states in Fig. \ref{f:bre1}, 
which corresponds to Figs. \ref{f:corner}(a) and \ref{f:corner}(c). 

\begin{figure}[htb]
\begin{center}
\begin{tabular}{cc}
\includegraphics[width=.47\linewidth]{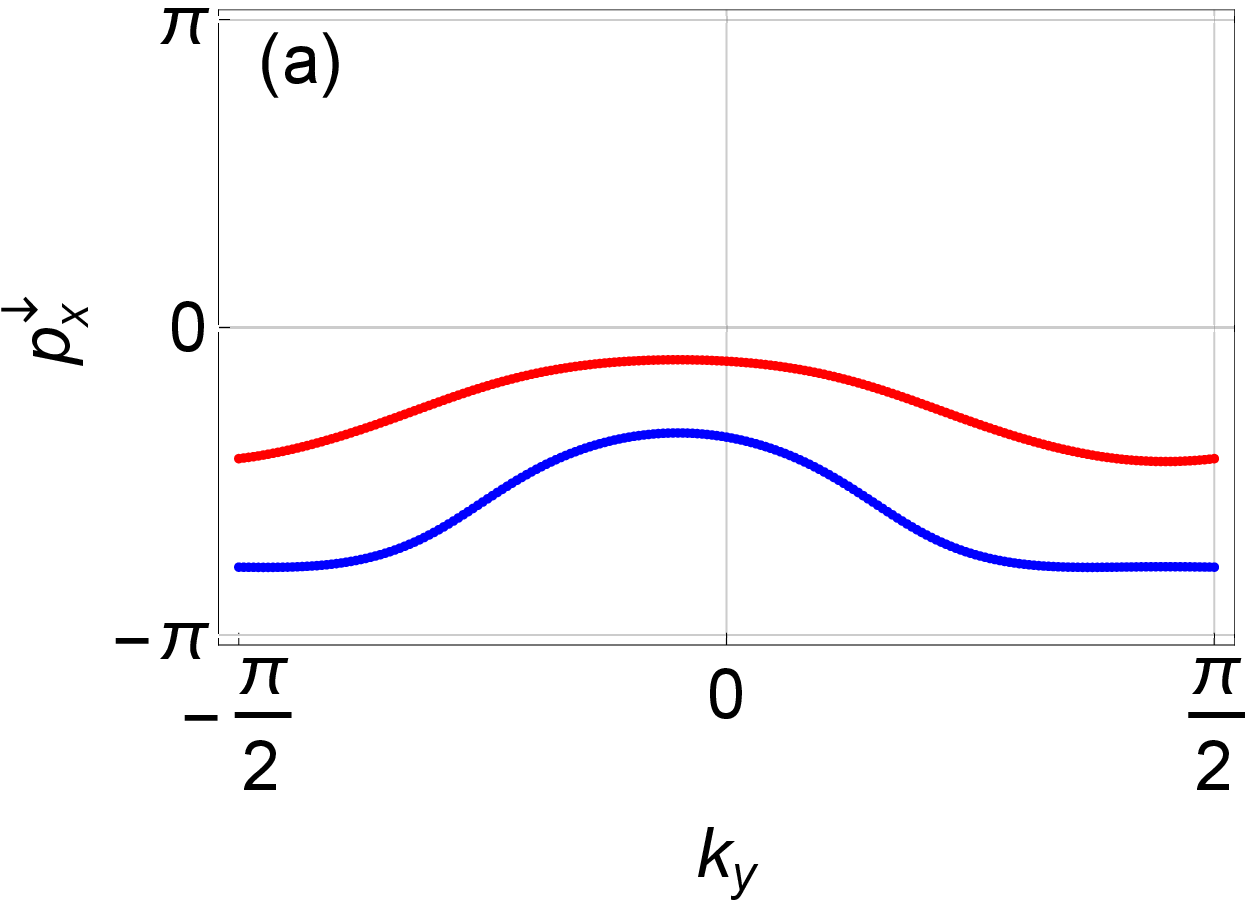}
&
\includegraphics[width=.52\linewidth]{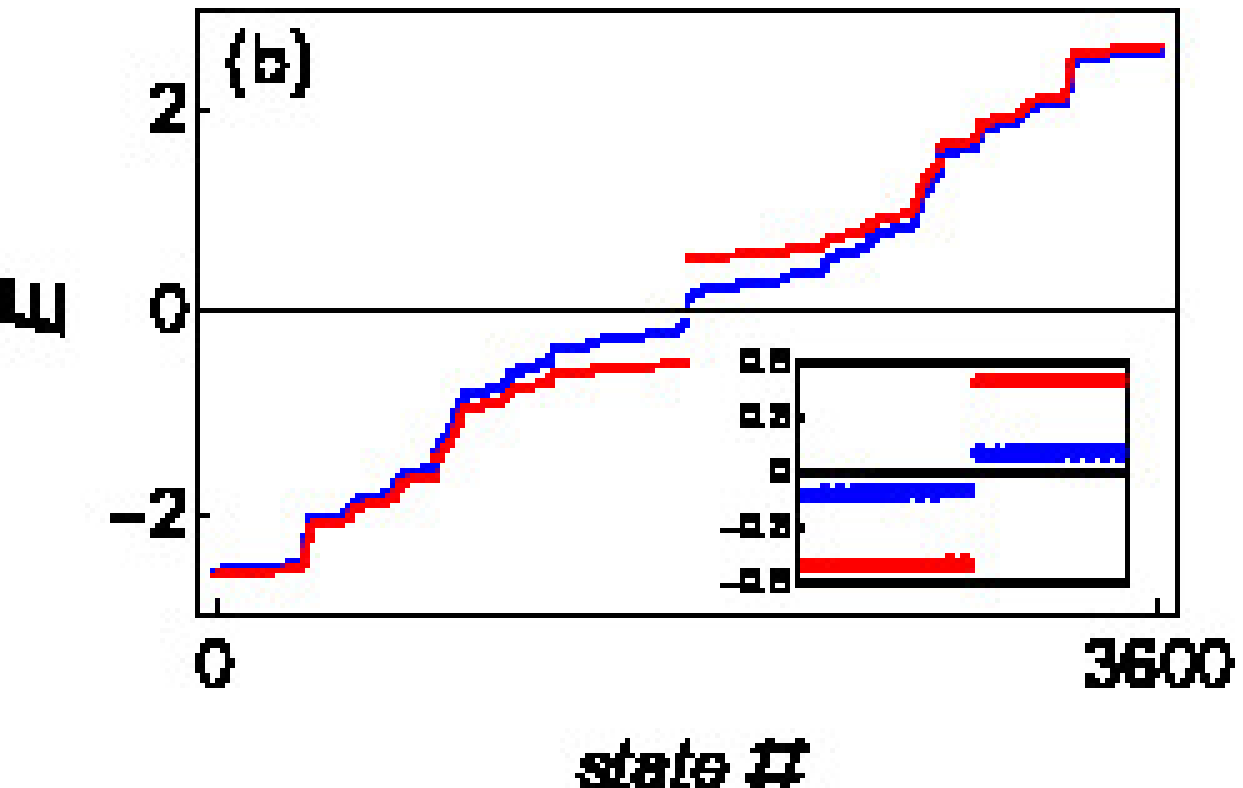}
\\
\includegraphics[width=.49\linewidth]{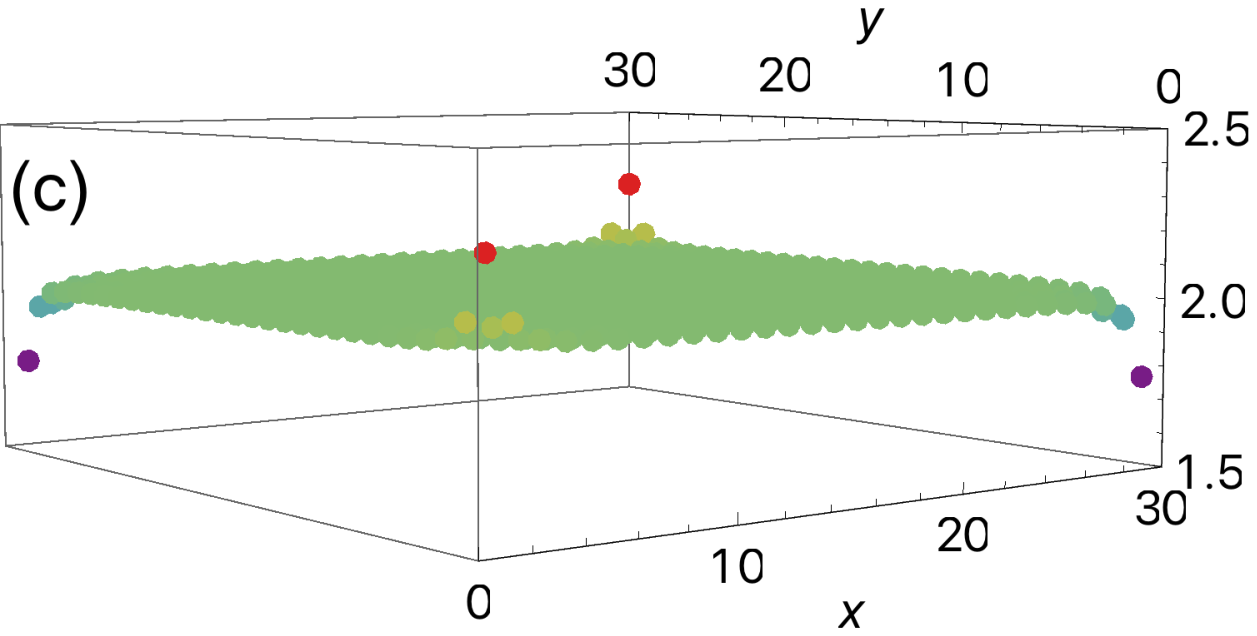}
&
\includegraphics[width=.49\linewidth]{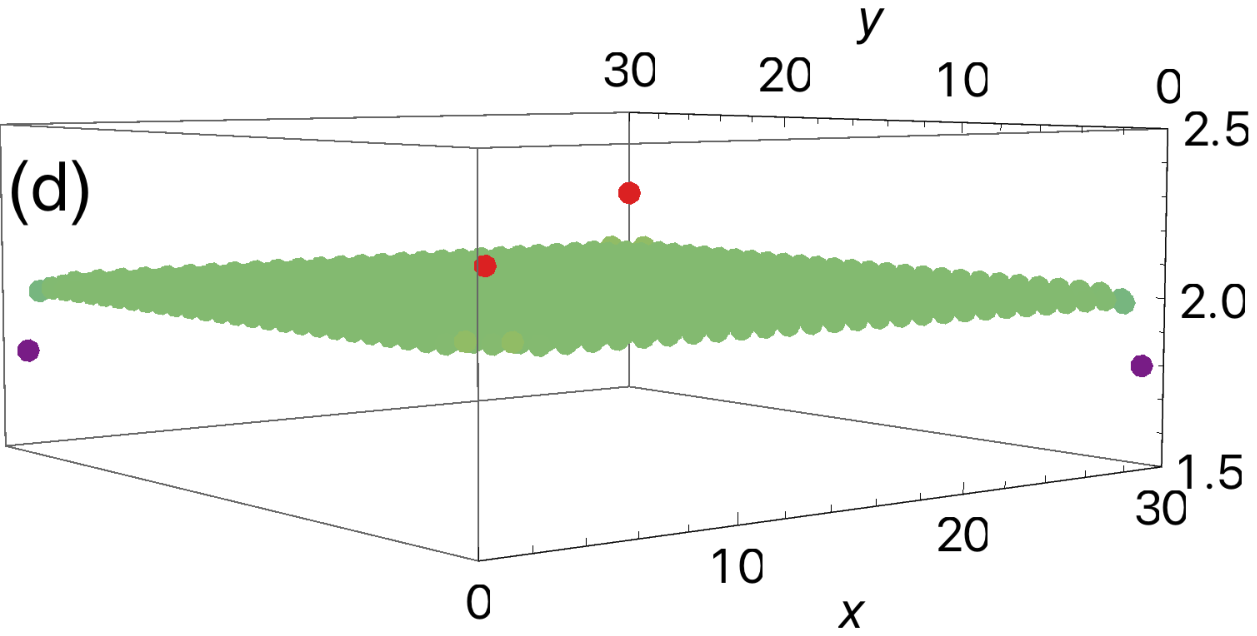}
\end{tabular}
\caption{
The same as Fig. \ref{f:bre1} but with $\phi=2\pi/10$, corresponding to Figs. \ref{f:corner}(b) and \ref{f:corner}(d).
}
\label{f:bre2}
\end{center}
\end{figure}

As already mentioned, the mother state in this case has eP $(p_x^\sigma,p_y^\tau)=(1/2,1/2)$. 
Such quantized eP change continuously if a finite $\delta$ in Eq. (\ref{Sta}) is introduced, as shown in Fig. \ref{f:bre1}(a).
Accordingly, the degenerate four zero energy states in the mother state in Fig. \ref{f:corner}(a) 
are lifted into two pairs of positive and negative energy states in Fig. \ref{f:bre1}(b).
The lifted pairs of zero energy mother states  still yield
the corner charges observed in Figs. \ref{f:bre1}(c) and \ref{f:bre1}(d).

This is also valid even if the gap of the mother states are very small. 
In Fig. \ref{f:bre2}, we show how the corner states change due to the finite staggered potentials for the mother state
in Figs. \ref{f:corner}(b) and \ref{f:corner}(d).
Since the gap of the mother state is small, lifted pairs of the zero energy states are soon absorbed into bulk spectrum, as in Fig. \ref{f:bre2}(b), 
even if $\delta$ is small. Therefore, we cannot recognize any signature of the zero energy states of the mother state in the spectrum.
Nevertheless, the corner charges survive even for a large $\delta=0.5 $ compared to the bulk gap $\sim0.1$,
as can be seen in Figs. \ref{f:bre2}(c) and \ref{f:bre2}(d).

\begin{figure}[htb]
\begin{center}
\begin{tabular}{cc}
\includegraphics[width=.47\linewidth]{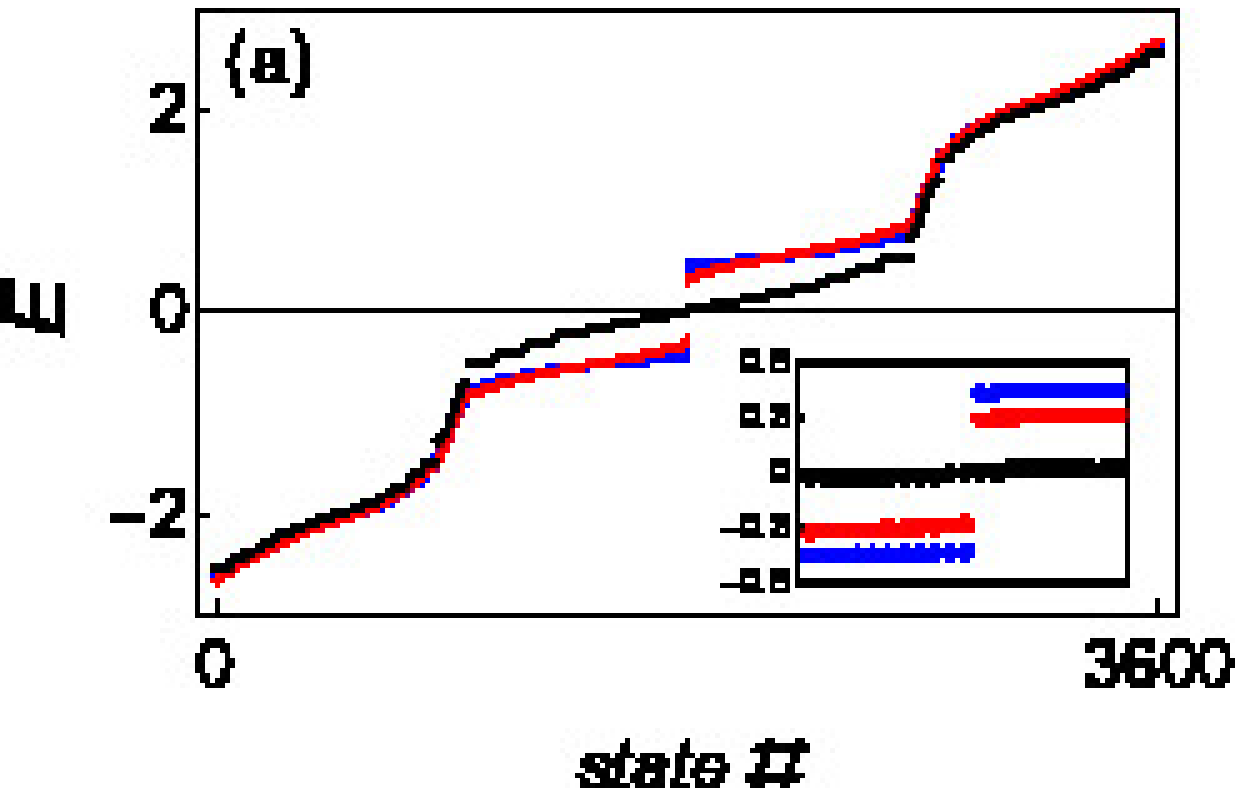}
&
\includegraphics[width=.49\linewidth]{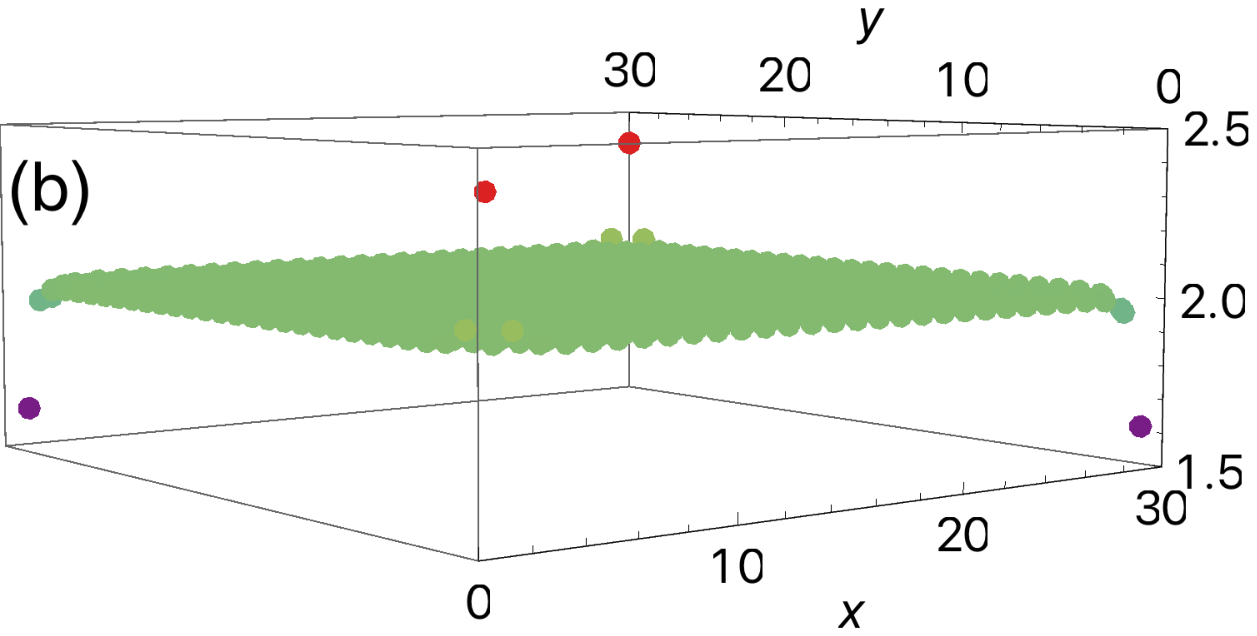}
\\
\includegraphics[width=.49\linewidth]{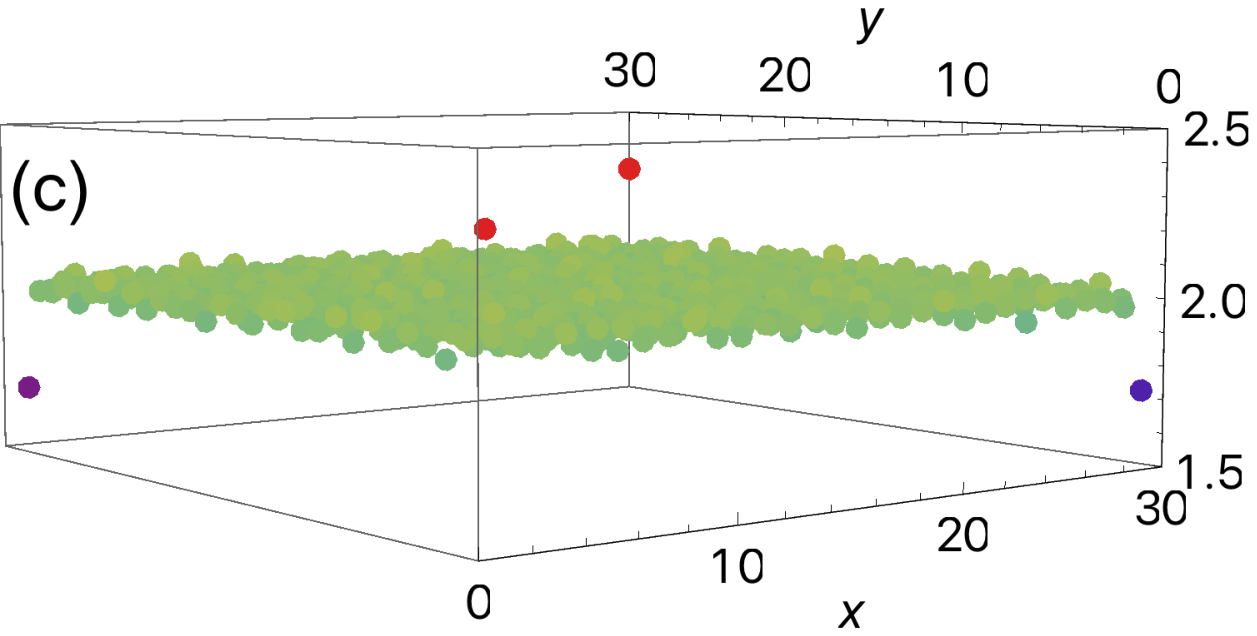}
&
\includegraphics[width=.49\linewidth]{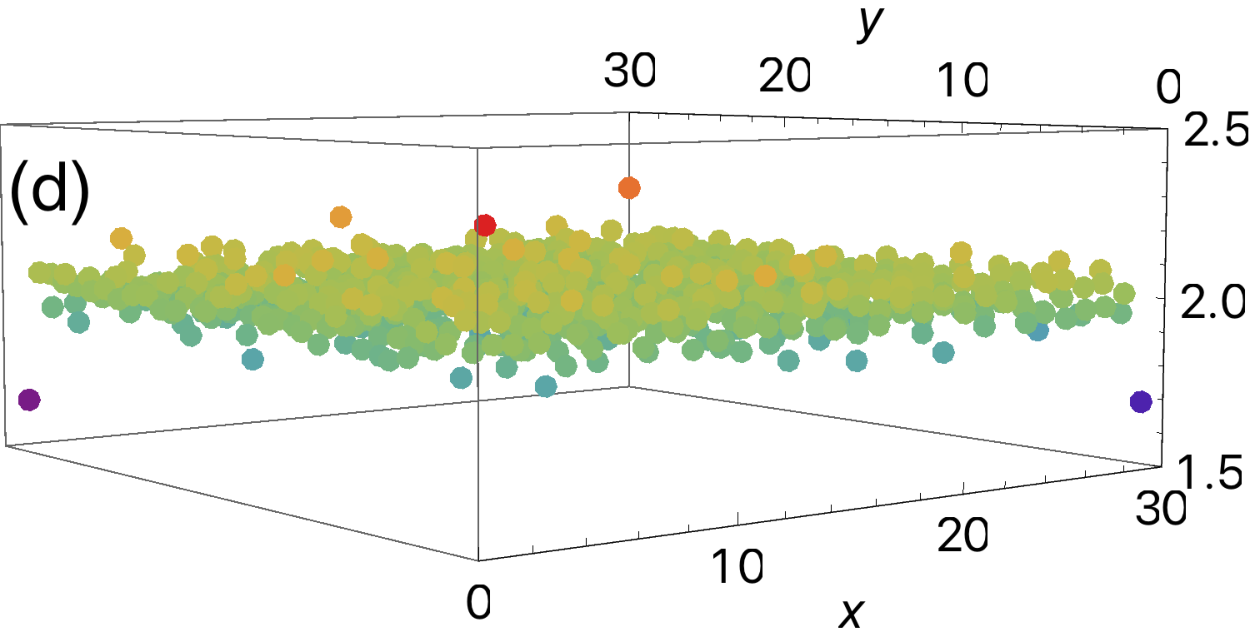}
\end{tabular}
\caption{
(a) Energies with full open boundary conditions for the model with
$\gamma=0.3$ and $\phi=2\pi/60$. Black dots show the energies of the mother state (i.e., with $\delta\sim0$ and 
$w=0$), whereas blue and
red ones include the staggered potential $\delta=0.5$ as well as
onsite disorder potential $\delta_j$ with $w=0.1$ and $w=0.3$, respectively. 
(b-d) show 
the occupied charge disributions for black, blue, and red cases above.
}
\label{f:bre3}
\end{center}
\end{figure}

Finally, we show the corner charges for a system with small $\gamma=0.3$, corresponding to the 
third panel from left in Fig. \ref{f:series_Hof}, with a rather weak magnetic field $\phi=2\pi/60$.
Although the gap of the mother state is extremely small, as seen in Fig. \ref{f:bre3}(a) (black dots), 
the corner states can be observed clearly in Fig. \ref{f:bre3}(b).
This holds true even if one introduces a large staggered potential yielding a large bulk gap similar to
the previous examples.
Once a large gap is open, the corner states are expected robust against disorder, although they are not protected by symmetries.
To exemplify  this, let us introduce an onsite disorder potential
\begin{alignat}1
H_{\rm d}=\sum_j\delta_jc_j^\dagger c_j,
\end{alignat}
where $\delta_j$ is a random real number restricted to $|\delta_j|<w$.
In Figs. \ref{f:bre3}(c) and \ref{f:bre3}(d), we show the distribution of the occupied charges in the SSH unit cells, 
in which one can distinguish corner charges from random bulk distribution, even for strong disorder 
$w\sim\gamma,\delta$.

This holds valid even for the critical model at $\phi=0$, in which the zero energy corner states are embedded in the bulk spectrum.
Therefore, we conclude that through the surviving corner states against the staggered potential,
one still has the possibility of 
observing a signature of the symmetry-protected corner states of the mother model.
We finally emphasize that even if the gap is extremely small in a weak field regime, 
the states belong to the same HOTI phase of the BBH state at least for a small $\gamma$.

\section{Summary and discussion}\label{s:sum}
In summary, we studied the 2D SSH model in a uniform magnetic field which interpolates simple SSH model and
the BBH model including $\pi$ flux. 
In other words, we extend the original Hofstadter butterfly 
by introducing the bond-alternation parameter $\gamma/\lambda$.
We showed that in such a generalized butterfly spectrum spanned by $\gamma/\lambda$ and $\phi$, 
the HOTI phase could exist in a rather wide region.

In the gapped region around $\pi$, the HOTI states belong to the topological quadrupole phase of the BBH model.
Therefore, if C$_4$ symmetry is broken, we expect that the models show the topological phase transition 
due to the gap-closing of the edge states.
Namely, the model would be in boundary-obstructed topological phases (BOTP) recently proposed by the authors of Ref. \cite{1908.00011}.
It may be quite interesting to investigate whether HOTI in other gapped regions in the butterfly spectrum
belong to the BOTP. 

\acknowledgements

This work was supported in part by Grants-in-Aid for Scientific Research Numbers 17K05563 and 17H06138
from the Japan Society for the Promotion of Science.
 
\appendix
\section{Entanglement polarization}\label{s:app}
In this Appendix, we discuss the definition and properties of eP introduced in Sec. \ref{s:ep} in detail.

\subsection{Entanglement polarization for the bulk}
As mentioned in Sec. \ref{s:ep}, for noninteracting systems, the eH also reduce to 
noninteracting  Hamiltonians \cite{Peschel:2003uq} written as,
$H^{A}=\sum_{j,l\in{A}}c_j^\dagger(k){\cal H}^{A}_{jl}(k)c_{l}(k)$,
where we assumed that the partition into $A$ and $\bar A$ keeps translational invariance.
The entanglement topological numbers are associated with the eigenfunctions $\chi^A(k)$ 
of the eH, obeying
\begin{alignat}1
\sum_{l}{\cal H}^{A}_{jl}(k)\chi^{A}_{ln}(k)=\chi^{A}_{jn}(k)\varepsilon^{A}_{n}(k),
\label{EntHam}
\end{alignat}
where $j,l\in A$.
Assume that the entanglement spectrum (eS), $\varepsilon^A_n(k)$,  is gapped at zero energy. 
Then, in the Schmidt-decomposition of the ground state
$|G\rangle=\sum_{a\in A,\bar b\in\bar A}D_{a\bar b}|\Psi_a\rangle\otimes|\Phi_{\bar b}\rangle
=\sum_j\lambda_j|\tilde\Psi_j\rangle\otimes|\tilde\Phi_j\rangle$,
the most dominant term is unique, given by  
$|G\rangle\sim|G^A\rangle\otimes|G^{\bar A}\rangle$, where $|G^A\rangle$ and
$|G^{\bar A}\rangle$ are the ground states of $H^A$ and $H^{\bar A}$, respectively \cite{Fukui:2015fk}.
Therefore, the topological number associated with $|G\rangle$ is just the sum of those associated with
$|G^A\rangle$ and $|G^{\bar A}\rangle$.
This motivates us to introduce the entanglement Berry connection (eBC),
\begin{alignat}1
A^{A}_{\mu}(k)=\sum_{\varepsilon^{A}_{n}<0}\sum_{j\in A}\chi^{A\dagger}_{nj}(k)\partial_{k_\mu}\chi^{A}_{jn}(k).
\end{alignat}
This is  the basis for various topological numbers associated with $|G^A\rangle$.
For the present model, the eP defined by
\begin{alignat}1
&p^{A}_{\mu}(k_\nu)=\frac{1}{2\pi i}\int_{-\pi_\mu}^{\pi_\mu} d k_\mu A^{A}_{\mu}(k),\quad (\mu\ne\nu),
\end{alignat}
where $\pi_x=\pi/q$ and $\pi_y=\pi/2$,
characterizes the HOTI and the corner states, as we shall show below.

It was pointed out \cite{Peschel:2003uq} 
that the eigenstates of the eH ${\cal H}^A(k)$ 
can be computed in a simpler way as follows:
Define the projection operator to the ground states,
\begin{alignat}1
P_{{\rm G},jl}(k)=\sum_{{\rm occ.~} n}\psi_{jn}(k)\psi_{nl}^\dagger(k),
\label{ProOpeG}
\end{alignat}
where $\psi_{jn}(k)$ is the eigenstate of ${\cal H}(k)$ in Eq. (\ref{OriHam}), obeying
$\sum_l{\cal H}_{jl}(k)\psi_{ln}(k)=\psi_{jn}(k)\varepsilon_n(k)$, and  the sum over $n$ 
is restricted to the occupied bands.
Note that $P_{{\rm G}}(k)$ obeys the same symmetry properties of ${\cal H}(k)$ in Eq. (\ref{Sym}).
Now, if $j,l$ in Eq. (\ref{ProOpeG}) are restricted to those belonging to $A$, which may be written as
\begin{alignat}1
P_{\rm G}^{A}(k)\equiv P^AP_{\rm G}(k)P^A,
\label{ProPro}
\end{alignat}
where $P^A$ stands for the projection 
operator to $A$, the eigenstates of $P_{\rm G}^{A}(k)$ is the simultaneous eigenstates of the 
eH, ${\cal H}^A(k)$ \cite{Peschel:2003uq}.
The eigenvalues of the eH, $\varepsilon^A(k)$ in Eq. (\ref{EntHam})
are related with those of the projection operator $\xi^A(k)$ in Eq. (\ref{ProPro})
as $\xi^A(k)=1/(e^{\varepsilon^A(k)}+1)$.
We therefore often call $P_{\rm G}^{A}(k)$ eH also.
As partitions, we consider two choices in Sec. \ref{s:ep}:
One is $A=\{1,2,\ldots,q\}\equiv\,\downarrow$ sites,
and the other is $A=\{1,3,\ldots,q-1,q+1,q+3,\ldots,2q-1\}\equiv{L(\rm eft)}$ sites in the magnetic unit cell in  Fig. \ref{f:lat}. 
Here, left means the left sites in the SSH unit cell.
Their complements are denoted as $\uparrow$ and $R$(ight), respectively.
In what follows, we often use $\sigma=\,\downarrow$ or $\uparrow$ and $\tau={L}$ or $R$, and $-\sigma$ and $-\tau$ stand for 
the complement of $\sigma$ and $\tau$, respectively. 
  
Finally, let us consider the symmetry properties of the eH.
The projection operators to $A=\sigma$ transforms as $\tilde M_xP^\sigma\tilde M_x^{-1}=P^{\sigma}$ and
$\tilde M_yP^\sigma\tilde M_y^{-1}=P^{-\sigma}$, whereas for the partition $A=\tau$,
as $\tilde M_xP^\tau\tilde M_x^{-1}=P^{-\tau}$ and
$\tilde M_yP^\tau\tilde M_y^{-1}=P^{\tau}$.
Therefore, we have
\begin{alignat}1
&\tilde M_yP_{\rm G}^\sigma(k_x,k_y)\tilde M_y^{-1}=P_{\rm G}^{-\sigma}(-k_x,k_y),
\nonumber\\
&\tilde M_xP_{\rm G}^\tau(k_x,k_y)\tilde M_x^{-1}=P_{\rm G}^{-\tau}(k_x,-k_y).
\end{alignat}
These symmetry properties give the following constraints on the eBC,
$A^{-\sigma}_x(-k_x,k_y)=A^{\sigma}_{x}(k_x,k_y)$ and 
$A^{-\tau}_{y}(k_x,-k_y)=A^{\tau}_{y}(k_x,k_y)$ appart from gauge transformations.
This leads to the following relationship, 
\begin{alignat}1
\begin{array}{l}
p^{\sigma}_{x}(k_y)=p^{-\sigma}_{x}(k_y)
\\
p^\tau_y(k_x)=p^{-\tau}_y(k_x)
\end{array}
+\mbox{(integer)},
\end{alignat}
where the integer is due to gauge ambiguities of eBC above.
On the other hand, 
the conventional polarizations of the half-filled ground states in Fig. \ref{f:hof} vanish 
for both directions $x$ and $y$,  implying 
$p^{\sigma}_x(k_y)+p^{-\sigma}_x(k_y)=p^{\tau}_y(k_x)+p^{-\tau}_y(k_x)=0$.
Therefore, $p^{\sigma}_x(k_y)$ and $p^\tau_y(k_y)$ 
should be quantized, taking only $0$ or $1/2$ modulo an integer.
Since the ground state for a fixed $\phi$ keeps a gap over the Brillouin zone, $p^\sigma_x$ 
and $p^\tau_y$ cannot depend on $k_y$ and $k_x$, respectively. 
Thus, the set of bulk eP, $(p^\sigma_x,p^\tau_y)$, can be topological invariants 
characterizing the HOTI.


\end{document}